\documentclass[prx,twocolumn,longbibliography]{revtex4}

\usepackage[utf8]{inputenc}
\usepackage{bm}
\usepackage{mathtools}
\usepackage{bbold}
\usepackage{amsmath}
\usepackage{amssymb}
\usepackage[colorlinks, linkcolor=black, anchorcolor=blue,citecolor=blue,urlcolor=blue]{hyperref}
\usepackage[normalem]{ulem}
\usepackage{color}
\usepackage{float}
\usepackage[dvipsnames]{xcolor}
\usepackage{physics}

\begin{document}

\title{Exponential onset of scalable entanglement via twist-and-turn dynamics in XY models}
\author{Tommaso Roscilde$^1$, Meenu Kumari$^{2,3,4}$, {Alexandre Cooper$^{4,5}$,} and Fabio Mezzacapo$^1$} 
\affiliation{$^1$Univ Lyon, Ens de Lyon, CNRS, Laboratoire de Physique, F-69342 Lyon, France}
\affiliation{{$^2$ Digital Technologies, National Research Council Canada}}
\affiliation{$^3$ Perimeter Institute for Theoretical Physics, Waterloo ON N2L 2Y5, Canada}
\affiliation{{$^4$ Institute for Quantum Computing, University of Waterloo, Ontario, Canada N2L 0A4}}
\affiliation{{$^5$ Department of Physics and Astronomy, University of Waterloo, Ontario, Canada N2L 3G1}}


\begin{abstract}
The efficient preparation of scalable multipartite entanglement is a central goal in the development of next-generation quantum devices. In this work, we show that the so-called ``twist-and-turn" (TaT) dynamics for interacting spin ensembles, generated by Hamiltonians with U(1)-symmetric interactions and with a transverse field, can offer an important resource to reach this goal. For models with sufficiently high connectivity, TaT dynamics exhibits two key features: 1) it features both scalable squeezing at short times, as well as quantum Fisher information with Heisenberg scaling at later times; and  2) scalable multipartite entanglement (up to Heisenberg scaling) is reached in a time growing only logarithmically with system size, associated with an exponential buildup of quantum correlations. These results can be shown exactly in the XY model with a Rabi field and infinite range interactions, and numerically in the case of spatially decaying XY interactions, such as dipolar interactions in two dimensions, provided that unstable spin-wave modes do not develop for large system sizes and/or strong fields. For dipolar interactions, the entanglement dynamics at intermediate times is completely at odds with thermalization; and it appears to saturate the maximum speed of entanglement buildup allowed by Lieb-Robinson bounds generalized to power-law interacting systems.   
\end{abstract}
\maketitle

\section{Introduction}
The dynamics of quantum information is a central subject of quantum many-body physics. It dictates fundamental aspects, such as:  whether and how quantum systems relax to thermal equilibrium \cite{dalessio_quantum_2016,Kaufmanetal2016,Abaninetal2019}; whether Hamiltonian evolution can lead to scrambling---{i.e.,} to the impossibility of recovering the initial state if a local perturbation acts on the system before the evolution is run backwards \cite{Garttner2017,Lietal2023,XuS2024}; and, most importantly for this work, the ultimate speed at which quantum devices, such as quantum computers and quantum simulators, can prepare and exploit entangled states \cite{Lucas_review}. A common paradigm in systems with local interactions---e.g., spin chains with short-range couplings---is that of a light-cone propagation of quantum information, or, equivalently, of a ballistic spreading of quantum correlations, regulated by so-called Lieb-Robinson bounds \cite{Lucas_review}. Within this picture, nearly free quasiparticles act as carriers of quantum information, propagating it at the maximum group velocity of their dispersion relation \cite{Cheneauetal2012,Carleoetal2014,Frerot2018PRL,Cevolanietal2019}. In contrast, systems in $D$ dimensions with interactions decaying as a power law of the distance $r$, {i.e.,} as $1/r^\alpha$, including the limit of long-range couplings ($\alpha < D$), can exhibit correlations that spread much faster than in systems with finite-range interactions. In particular, for $\alpha < 2D$, the spreading can be exponentially fast, {i.e.,} over distances growing as $\exp(t)$ with time $t$.  This behavior was first predicted as a generalization of Lieb-Robinson bounds to systems with power-law interactions \cite{HastingsK2006, Tran2021_b}, and it has been confirmed through the explicit design of protocols generating GHZ-like states using time-dependent power-law interactions \cite{Tran2021}. 
These observations naturally raise the question of whether such an extraordinary fast propagation of quantum correlations can be obtained as well for time-independent Hamiltonians, of relevance to the ongoing effort of quantum simulation of non-equilibrium quantum dynamics \cite{Polkovnikov2011,dalessio_quantum_2016,Abaninetal2019}. 

In this work, we show that a broad class of time-independent Hamiltonians for spin lattices can exhibit an exponential onset of entanglement. Multipartite entanglement emerges from dynamics fundamentally driven by zero-momentum nonlinear excitations (as opposed to finite-momentum quasiparticles). The dynamics under consideration is the so-called twist-and-turn (TaT) dynamics \cite{Michelietal2003,Muesseletal2015,Sorellietal2019,MunozArias2023}, which has so far been studied only for infinite-range interactions, and which we here generalize to spatially decaying interactions. TaT dynamics is generated by. In the following we shall denote as TaT Hamiltonian, $H_{\rm TaT}$, a Hamiltonian comprising XY interactions among spins that decay as a power law of the distance, together with an applied Rabi field:
\begin{equation}
H_{\rm TaT} = - \frac{\cal J}{{\cal N}_\alpha} \sum_{i\neq j} \frac{1}{r_{ij}^\alpha} (S_i^x S_j^x + S_i^y S_j^y) + \Omega \sum_i S_i^x,
\label{e.TaT}
\end{equation}  
where $S^\mu_i$ ($\mu = x,y,z$) are spin-1/2 operators attached to the sites $i$ of a lattice, which we take to be a $L \times L$ square lattice ($D=2$, $N=L^2$) with periodic boundary conditions. We choose the coupling to be ferromagnetic (${\cal J}>0$), and define the Kac factor as ${\cal N}_\alpha=1+ \sum_{r \neq 0} r^{-\alpha}$ for $\alpha \leq D$, and ${\cal N}_\alpha=1$ otherwise. $\Omega$ is a transverse field chosen to point \emph{opposite} to the orientation of an initial coherent spin state $|\psi(0)\rangle = \otimes |\rightarrow_x\rangle_i$. 
The Hamiltonian $H_{\rm TaT}$ with infinite-range interactions ({i.e.,} $\alpha=0$)  is known to drive exponentially fast dynamics of entanglement  \cite{Michelietal2003, MunozArias2023}, and it has been implemented experimentally in spinor Bose-Einstein condensates \cite{Strobeletal2014, Muesseletal2015} as well as in cold gases coupled via a cavity \cite{Lietal2023}. Yet the scaling behavior of the entanglement properties of the evolved state has not been investigated theoretically, nor experimentally observed.   
Making use of analytical and numerically exact results, as well as of different cross-checked approximations on large lattices, we show that $H_{\rm TaT}$ exhibits exponentially fast scalable entanglement even for power-law interactions, {i.e.,} $\alpha>0$. In particular, we focus our attention on the case of dipolar interactions ($\alpha=3$) on square lattices, of immediate relevance to experiments using e.g., arrays of Rydberg atoms \cite{BrowaeysL2020}, magnetic atoms \cite{Chomazetal2022}, or dipolar molecules \cite{Cornish2024}.  Remarkably, the entanglement generated in the evolution is potentially relevant for metrological applications, both at short times (in the form of squeezing), as well as at longer ones \cite{FrerotR2024}. The longer-term entanglement reaches Heisenberg scaling exponentially in time over a finite range of sizes, which diverges when the transverse Rabi field vanishes. This scaling, corresponding to long-range entanglement, is completely at odds with the thermodynamics of the system, which is trivialized by the applied field. Hence the entangling dynamics cannot at all be understood as a thermalization process. 

The structure of the paper is as follows. We first discuss the squeezing dynamics and scaling properties of the TaT model with all-to-all interactions (Sec.~\ref{s.squeezing}). We then report the observation of scalable spin squeezing in the case of dipolar interactions (Sec.~\ref{s.dipsqueezing}) and describe the emergence of unstable modes in the spin-wave spectrum at finite wavevector (Sec.~\ref{s.SW}). We finally explore the onset of multipartite entanglement with Heisenberg scaling at later times (Sec.~\ref{s.Heisenberg}), before offering a summary of our findings (Sec.~\ref{s.conclusions}).


\begin{figure}[ht!]
\begin{center}
\includegraphics[width=\columnwidth]{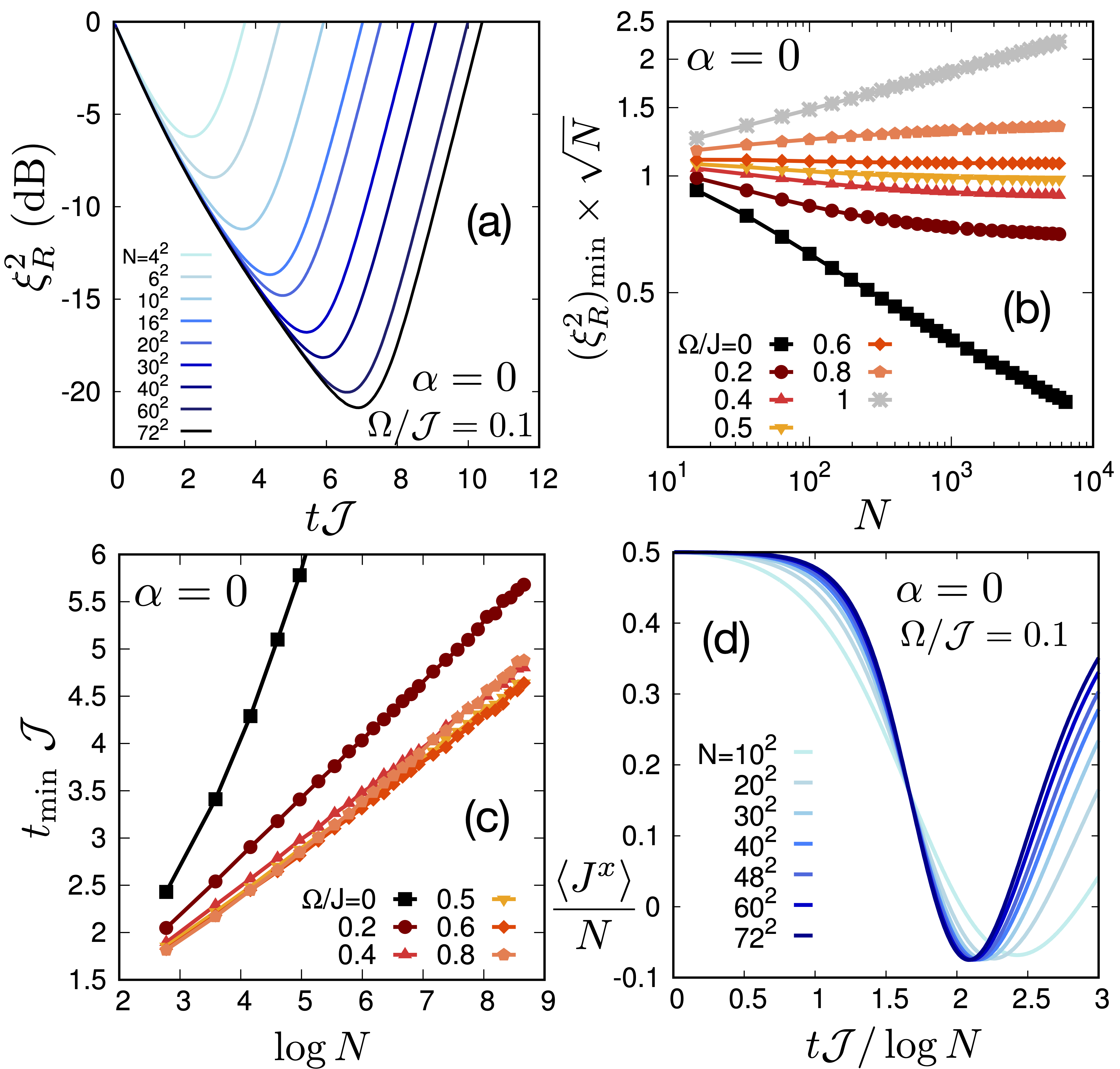}
\caption{\emph{Twist-and-turn squeezing dynamics with all-to-all interactions.} (a) Spin squeezing dynamics for $\Omega = 0.1 {\cal J}$;  (b) Optimal squeezing  and (c) time to optimal squeezing for the TaT dynamics with all-to-all  interactions; (d) Magnetization dynamics for  $\Omega = 0.1 {\cal J}$. All data stem from exact calculations. }
\label{f.squeezinga0}
\end{center}
\end{figure}

\begin{figure*}[ht!]
\begin{center}
\includegraphics[width=\textwidth]{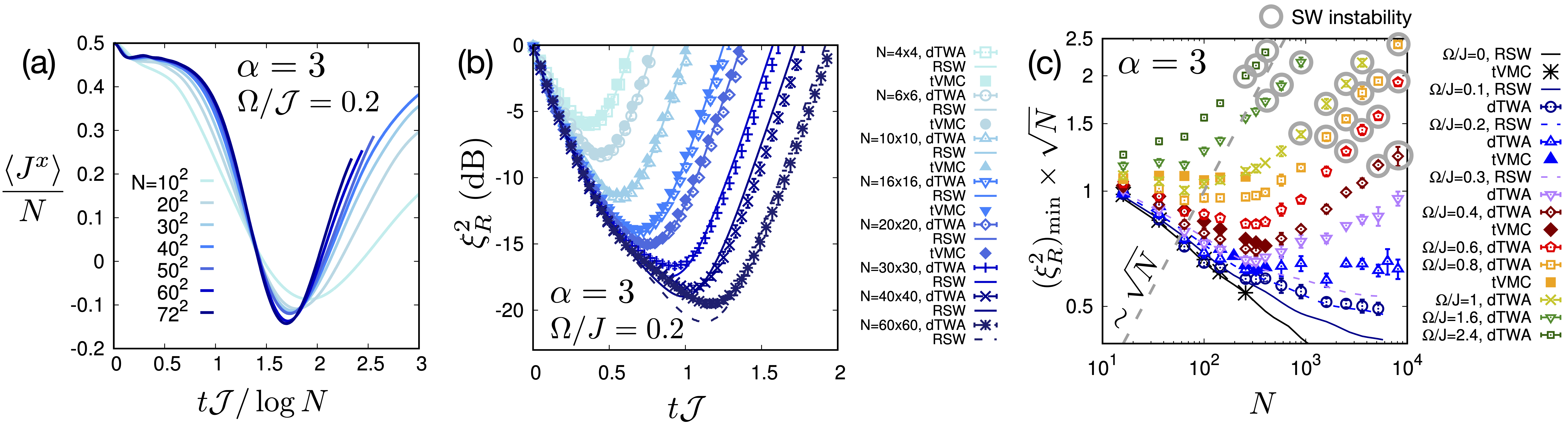}
\caption{\emph{Twist-and-turn squeezing dynamics with dipolar interactions.} (a) Dynamics of the magnetization of the 2D dipolar model with $\Omega/{\cal J} = 0.2 $ (dTWA data). (b) Squeezing dynamics for the dipolar 2D system with $\Omega/{\cal J}=0.2$, from dTWA, RSW and tVMC. (c) Scaling of the optimal squeezing for different values of the field. The dashed grey line indicates the $\sqrt{N}$ behavior, corresponding to the absence of scalable squeezing. The grey circles highlight field and size values for which imaginary spin-wave (SW) frequencies have appeared.}
\label{f.squeezinga3}
\end{center}
\end{figure*}

\section{Twist-and-turn dynamics leading to scalable squeezing} 
\label{s.squeezing}

For systems with all-to-all interactions ($\alpha=0$), $H_{\rm TaT}$ is well known to exhibit squeezing at short times, as detected by the squeezing parameter \cite{Pezze2018RMP}
\begin{equation}
\xi_R^2 = \frac{N \min_\perp {{\rm Var}(J^\perp)} }{\langle J^x \rangle^2}
\end{equation}
where ${\bm J} = \sum_i {\bm S}_i$ is the collective spin operator; and the numerator is minimized over the collective spin components perpendicular to the net magnetization appearing along the $x$ axis. The uncertainty region on the Bloch sphere of the initial, coherent spin state is rotationally symmetric around the $x$ axis; however, it gets deformed by the dynamics generated by $H_{\rm TaT}$, with the spin-spin interaction term leading to ``twisting" of the collective spin ({i.e.,} to a rotation around the $z$ axis in the northern hemisphere, or around the $-z$ axis in the southern hemisphere), and the transverse field term leading to ``turning" of the spin around the $x$ axis \cite{Muesseletal2015}. As discussed in Ref.~\cite{MunozArias2023}, in the classical limit of a collective spin of infinite length, and for $0 < \Omega <  {\cal J}$, the positive $x$ direction represents a hyperbolic fixed point. Its directions of stability and instability correspond to exponential squeezing (see Fig.~\ref{f.squeezinga0}(a)) and anti-squeezing, respectively, with a rate $\lambda = \sqrt{\Omega({\cal J} -\Omega)}$. 

At the quantum mechanical level, exponential squeezing and anti-squeezing at short times with the same exponent $\lambda$ can be obtained by mapping the collective spin (of fixed length $J = N/2$) onto a bosonic mode---see App.~\ref{a.bosonic} for a detailed derivation---namely $\min_\perp {{\rm Var}(J^\perp)} \approx \frac{N}{4} e^{-2\lambda t}$. The fastest squeezing dynamics is therefore achieved for the optimal field $\Omega_o/{\cal J}  =1/2$. Nonetheless squeezing dynamics is found to stop at times $t_{\rm min} \sim \frac{1}{\lambda} \log \sqrt{N}$ (Fig.~\ref{f.squeezinga0}(a,c)) \cite{MunozArias2023}, and to lead therefore to an optimal squeezing scaling as $(\xi_R^2)_{\rm opt} \sim 1/\sqrt{N}$. This behavior, revealed through exact diagonalization, is found to be universal across the entire field interval $0 < \Omega < {\cal J}$, as shown in Fig.~\ref{f.squeezinga0}(b). 
Compared with one-axis-twisting (OAT) dynamics (corresponding to the case $\alpha=0$, $\Omega=0$), TaT squeezing dynamics leads to a weaker scaling of optimal squeezing: $\xi_R^2 \sim N^{-1/2}$ for TaT vs $\xi_R^2 \sim N^{-2/3}$ for OAT. However, TaT dynamics has the advantage that this optimal squeezing is reached in a time that grows only logarithmically with system size, due to an exponential buildup of spin squeezing at short times. 

\section{Dipolar twist-and-turn squeezing} 
\label{s.dipsqueezing}
We now consider power-law decaying interactions, focusing on the case of dipolar interactions ($\alpha=3$) in two dimensions. Such interactions allow the dynamics to leak out of the Dicke sector of maximal spin length $\langle {\bm J}^2 \rangle = N/2(N/2+1)$; and, at least in principle, to relax toward states that reproduce locally the behavior of a thermal state at the energy of the initial state. Thermalization under $H_{\rm TaT}$ would require the magnetization to reverse its orientation (to align with the field), and the excess Rabi energy to be dumped into the interaction term, thereby suppressing correlations (more on this aspect below). 
 
Yet, this is not at all what is observed. To reconstruct the dynamics of the dipolar system, we employ a joint numerical strategy based on the discrete truncated Wigner approximation (dTWA) \cite{Schachenmayer2015PRX} and time-dependent variational Monte Carlo using the pair-product Ansatz \cite{PRB2019, Comparin2022PRA}, augmented with a wavefunction term that depends on the total magnetization \cite{Calecaetal2024}. For sufficiently small fields, the magnetization dynamics of the dipolar system remains close to that of an all-to-all coupled system (Fig.~\ref{f.squeezinga0}(d)), but with an effective coupling constant ${\cal J}_{\rm eff} = {\cal J} \sum_{r\neq 0} r^{-\alpha}$ -- i.e., with an interaction term of the kind $({\cal J}_{\rm eff}/N) [(J^x)^2 + (J^y)^2] $. And the magnetization partially reverses direction over a time scaling as $\log N$---see Fig.~\ref{f.squeezinga3}(a). 
This dynamics leaves ample time for squeezing correlations to build up within a timescale growing also logarithmically with size, and for optimal squeezing to reach the same scaling as in the $\alpha=0$ case. In Figs.~\ref{f.squeezinga3}(b-c), we clearly observe this behavior for $\Omega/{\cal J} = 0.2$ using both numerical approaches and up to very large system sizes ($L=90$ using dTWA). 
 
This behavior is quantitatively captured by a rotor-spin-wave (RSW) separation scheme \cite{Roscildeetal2023}, which describes the dynamics in the Dicke sector exactly as that of the $\alpha=0$ model with ${\cal J}_{\rm eff}$ coupling---and treats the leakage of the dynamics from the Dicke sector within a spin-wave approximation---see App.~\ref{a.RSW} for an extended discussion. RSW theory deviates from dTWA for very large sizes ($N \gtrsim 30^2$), as the two theories predict a change in the exponential decay of squeezing below -15 dB, but with a different exponent (see Fig.~\ref{f.squeezinga3}(b)). 

A comprehensive picture of the scaling of the optimal squeezing for various fields is provided in Fig.~\ref{f.squeezinga3}(c). There we observe that, for sufficiently small fields ($\Omega/{\cal J} \lesssim 0.2$), optimal squeezing appears to maintain the same asymptotic scaling as in the $\alpha=0$ case, namely $(\xi_R^2)_{\rm min} \sim N^{-\nu}$ with $\nu = 1/2$. For larger fields, optimal squeezing remains scalable, but its asymptotic scaling appears to become field dependent, with a $\nu$ exponent decreasing gradually with $\Omega$.  Determining the field at which scalability is entirely lost is rather difficult. Nonetheless it is clear that scalable spin squeezing is observed in the dipolar dynamics, in spite of the emergence of unstable spin-wave modes, as we shall discuss in the next section.

\begin{figure*}[ht!]
\begin{center}
\includegraphics[width=\textwidth]{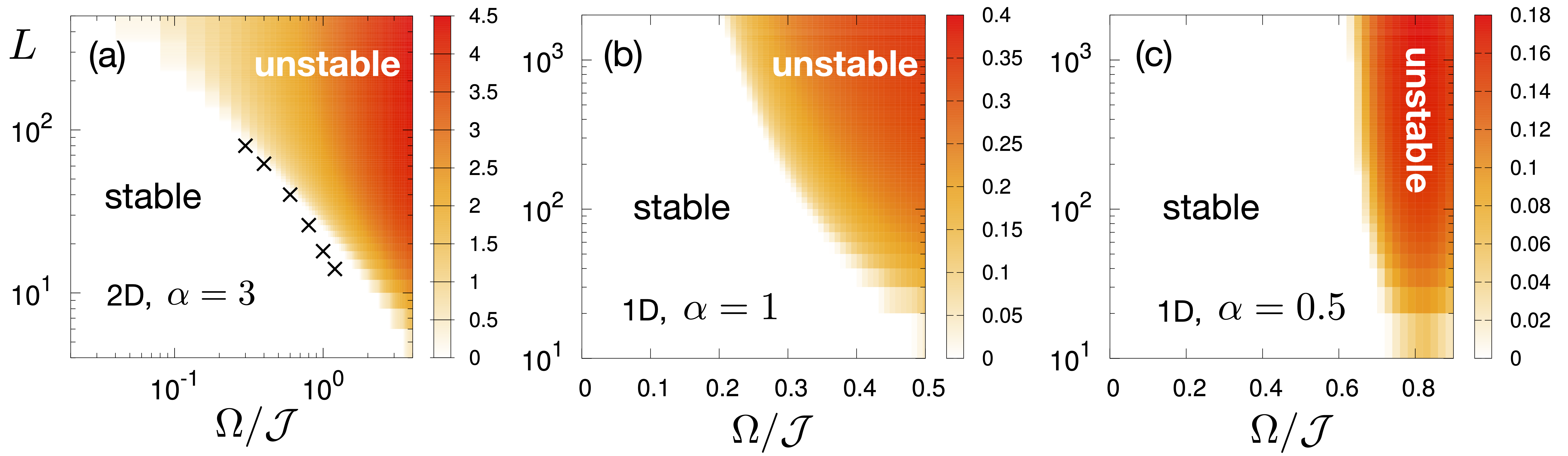}
\caption{\emph{Stability diagram of spin waves.} (a) Dipolar interactions ($\alpha = 3$) in 2D. The false colors indicate the largest imaginary spin-wave frequency at finite wavevector $\lambda_{\max} = \max_{\bm k \neq 0} {\rm Im}(\omega_{\bm k})$, as a function of the applied field $\Omega$ and the linear system size $L$. The crosses indicate the estimated crossover in the scaling of ${\rm Var}(J^y)_{\rm max}$---see Fig.~\ref{f.VarJydipscaling}. (b) Coulomb interactions ($\alpha = 1$) in 1D; (c) Long-range interactions ($\alpha = 0.5$) in 1D. Same significance of false colors as in panel (a). The difference in field ranges and values of the exponent between panel (a) and panels (b-c) should be mainly attributed to the different Kac normalization (see main text).}
\label{f.stability}
\end{center}
\end{figure*}

\section{Spin-wave instabilities}
\label{s.SW}
A peculiar feature of RSW theory applied to TaT dynamics is that it is built around an \emph{unstable} fixed point of the classical dynamics \cite{MunozArias2023}. In systems with spatially decaying interactions, the same instability that affects the dynamics of the zero-momentum degrees of freedom, leading to its exponential acceleration, can also affect that of the finite-momentum ({i.e.,} spin-wave) modes. Indeed, for sufficiently large sizes, the spin-wave (SW) frequencies of the dipolar XY model at small wavevector ($|\bm k| \sim 1/L$) can become imaginary $\omega_{\bm k} = i \lambda_{\bm k}$ ---see App.~\ref{a.RSW} for a detailed discussion. 

Fig.~\ref{f.stability}(a) shows the range of fields and sizes at which unstable SW modes appear for the 2D dipolar case. 
As discussed in App.~\ref{a.RSW}, an instability at finite momentum occurs for a field value which pushes the frequency of the mode at the lowest wavevector (${\bm k} = (2\pi/L,0)$ on a square lattice) to imaginary values. This condition is met when the field exceeds a critical value
\begin{equation}
 \left ( \frac{\Omega}{\cal J} \right)_c(L) ~ \sim L^{-2z}  ~~~~~ (L \gg 1)
 \label{e.Omegac}
\end{equation}
where $z$ is the dynamical exponent dictating the dispersion relation at zero field of the system at long wavelength, $\omega \sim k^z$. For the models of interest to this work, $z = 1$ for $\alpha \geq D+2$, $z = (\alpha-D)/2$ for $D < \alpha < D+2$, and $z = 0$ for $\alpha \leq  D$ \cite{Frerot2017PRB}. 

For the moderate fields on which we focus our discussion, the instability in the modes of the dipolar system occurs for very large system sizes---{e.g.,} above $N \approx 10^2\times 10^2$ for $\Omega/{\cal J} = 0.2$. Interestingly, Fig.~\ref{f.squeezinga3}(c)  shows that scalable optimal squeezing persists even for sizes and fields at which the SW spectrum develops unstable modes. This aspect can be related to the fact that the exponential rate of proliferation of low-$k$ spin waves, $\lambda_{\bm k}$, remains systematically smaller than the rate $\lambda$  of increase of fluctuations of the zero-mode fluctuations, hence affecting only mildly the early dynamics of the system. The dTWA data of  Fig.~\ref{f.squeezinga3}(c) suggests a clear crossover in the scaling properties for intermediate sizes, which could be interpreted as a signature of the onset of unstable modes. Yet, as we show in App.~\ref{a.Jx}, at the optimal squeezing time no clear dynamical instability is observed for the system sizes we investigated, since the collective spin remains firmly polarized $\langle J^x \rangle \sim O(N)$. The change in scaling of optimal squeezing is rather due to  $\min_\perp {{\rm Var}(J^\perp)}$ changing its exponentially decaying dynamics (as seen in Fig.~\ref{f.squeezinga3}(b) for the largest system sizes).  As we shall see below, unstable modes have instead a clearer impact on the later dynamics of entanglement.   

It is important to point out that, for $\alpha < D$, the critical field loses its size dependence in Eq.~\eqref{e.Omegac}, namely there is a whole range of field values for which spin waves built around an unstable point are actually stable harmonic modes for all system sizes. Fig.~\ref{f.stability}(b-c) shows the stability diagram for interactions with $\alpha = 1$ (Coulomb) and $\alpha = 0.5$ in 1D, both relevant to trapped ion chains \cite{Monroe2021RMP}. The Coulomb case is marginal, i.e., it sits at the boundary between short-ranged interactions and long-ranged ones, and it shows a logarithmic dependence of the critical field on the system size. On the other hand, in the case $\alpha=0.5$ in $D=1$, corresponding to long-range interactions, the size dependence of the critical field essentially disappears. This implies that there exists an entire class of models with power-law decaying interactions for which the TaT dynamics of the $\alpha=0$ limit is potentially stable for all system sizes.

\begin{figure*}[ht!]
\begin{center}
\includegraphics[width=\textwidth]{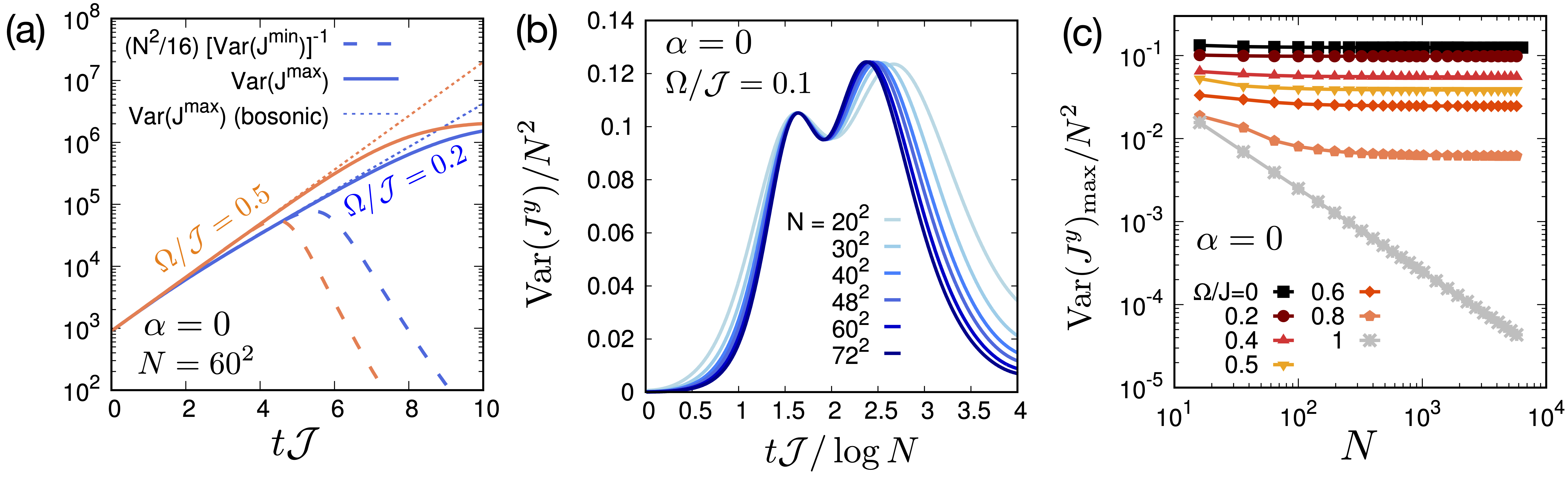}
\caption{\emph{Twist-and-turn exponential buildup of long-range entanglement.} (a) Exponential dynamics of the variances of the squeezed ($J^{(\min)}$) and anti-squeezed ($J^{(\max)}$) collective-spin component for $\alpha=0$ and two field values. The dashed line corresponds to the squeezing dynamics of the bosonic model obtained by linearized Holstein-Primakoff mapping of the collective spin;  (b) Dynamics of ${\rm Var}(J^y)$ for $\Omega = 0.1 {\cal J}$; (c) Scaling of the peak of ${\rm Var}(J^y)$ for $\alpha=0$. The value corresponds to the second local maximum for $\Omega>0$ (see panel b), and to the plateau value for $\Omega = 0$.}
\label{f.VarJyTaT}
\end{center}
\end{figure*}

\begin{figure}[ht!]
\begin{center}
\includegraphics[width=\columnwidth]{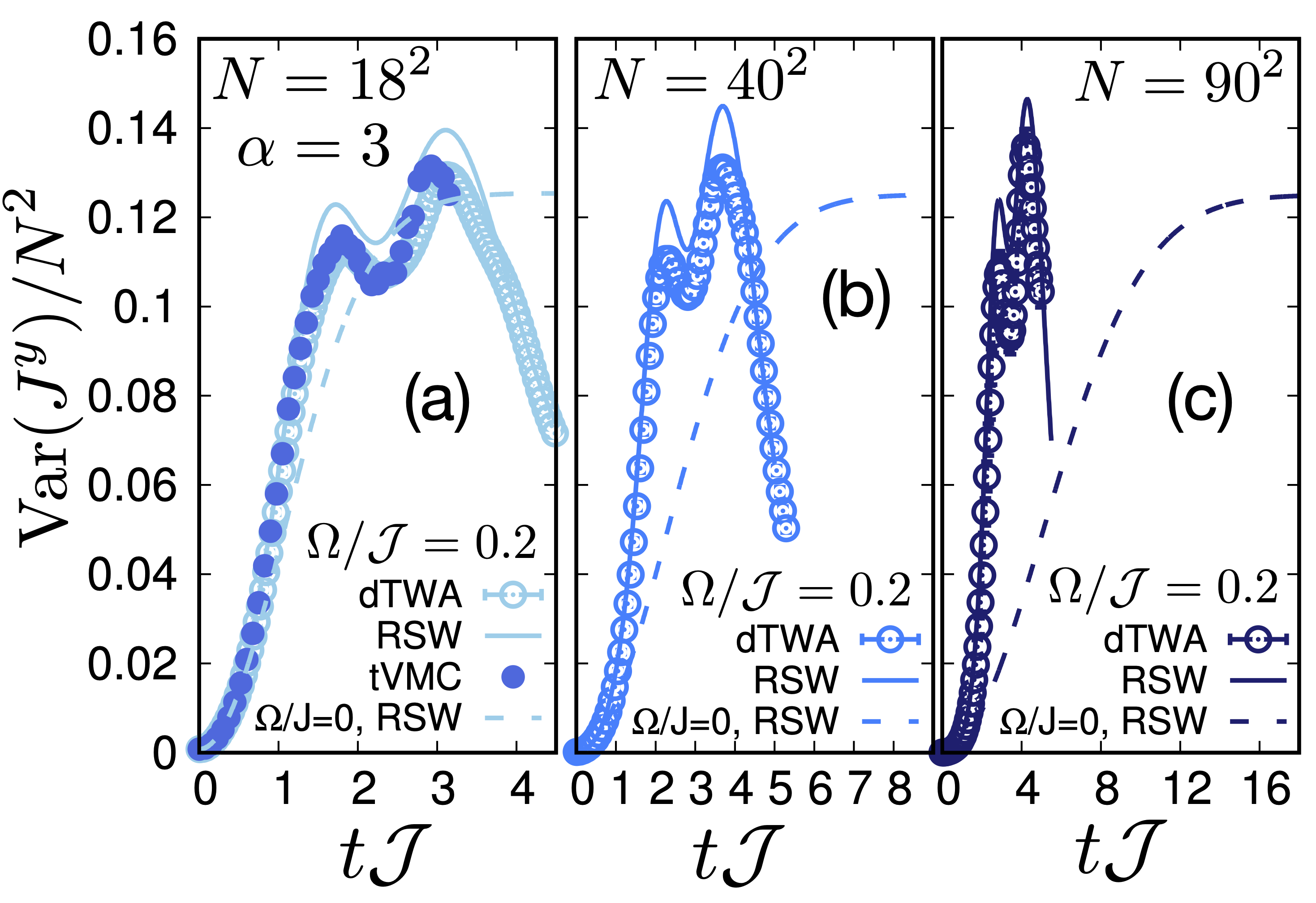}
\caption{\emph{Exponential onset of entanglement in the 2D dipolar model.} Dynamics of ${\rm Var}(J^y)$ for the 2D dipolar model with $\Omega/{\cal J} = 0.2$ from RSW, dTWA and tVMC, for three different system sizes. The OAT case ($\Omega = 0$) is also shown for comparison.}
\label{f.VarJydip}
\end{center}
\end{figure}

\begin{figure*}[ht!]
\begin{center}
\includegraphics[width=\textwidth]{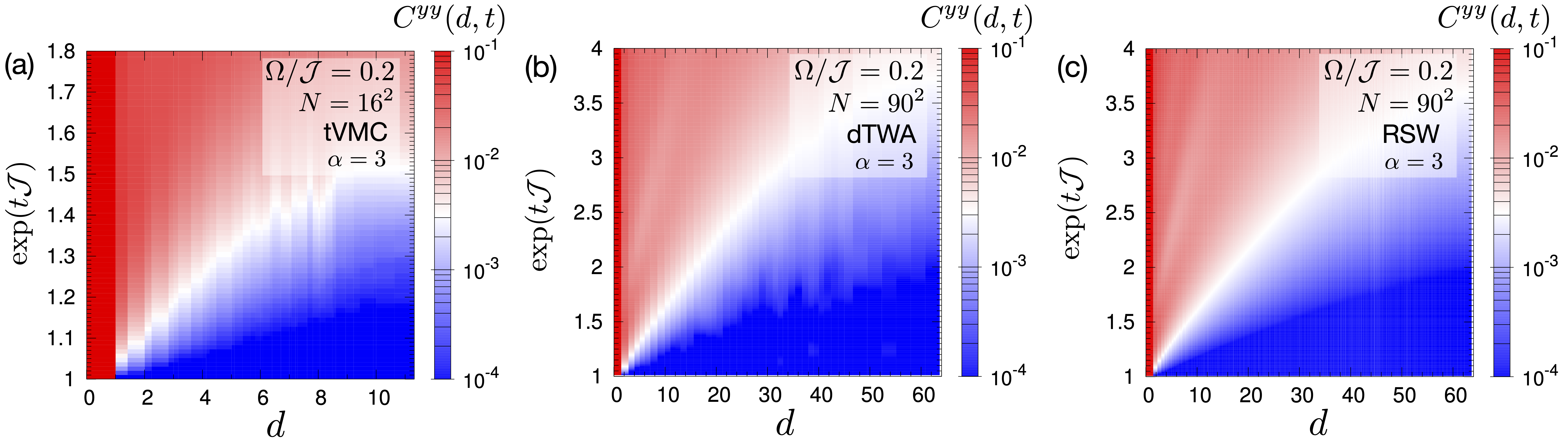}
\caption{\emph{Exponential spreading of correlations.}  Dynamics of transverse correlations $C^{yy}(d,t) = \langle S^y_i S^y_{i+d} \rangle(t)$ for the 2D dipolar model with $\Omega/{\cal J} = 0.2$ for $N=16^2$ from tVMC (a) and for $N=90^2$ from dTWA (b) and RSW (c).}
\label{f.correlations}
\end{center}
\end{figure*}

\begin{figure}[ht!]
\begin{center}
\includegraphics[width=0.9\columnwidth]{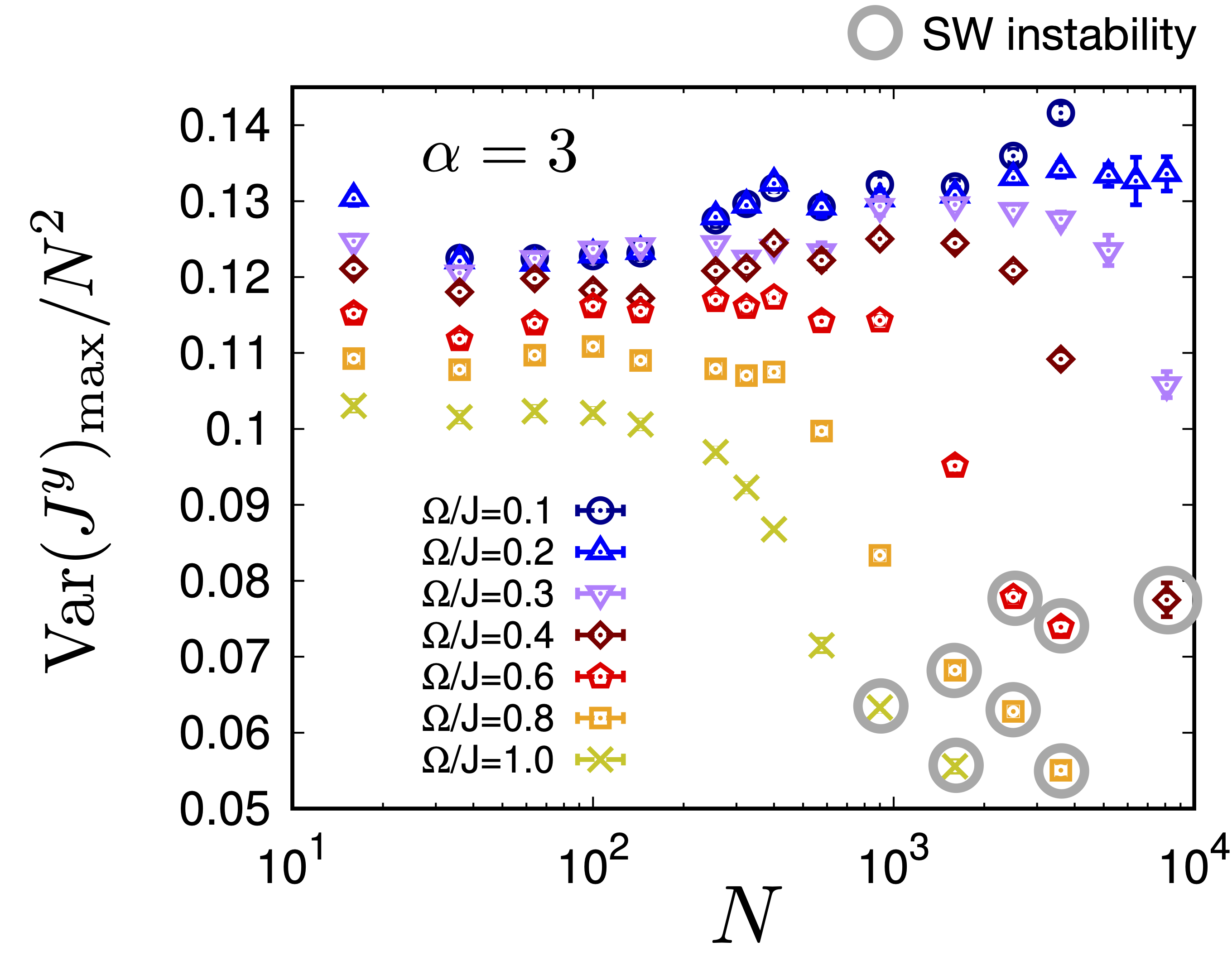}
\caption{\emph{Heisenberg scaling of entanglement and its breakdown.} Scaling of the peak of ${\rm Var}(J^y)$ for $\alpha=3$. As in Fig.~\ref{f.squeezinga3}, the grey circles highlight field and size values at which imaginary spin-wave (SW) frequencies appear.}
\label{f.VarJydipscaling}
\end{center}
\end{figure}

\section{Twist-and-turn dynamics leading to Heisenberg scaling of entanglement}
\label{s.Heisenberg}
 
 \subsection{All-to-all interactions}
We now focus our attention on the entangling dynamics after squeezing is lost.  As seen in Sec.~\ref{s.squeezing}, the exponential squeezing of the minimum variance stops at a time $\sim\log(\sqrt{N})$. Yet, as shown in Fig.~\ref{f.VarJyTaT}(a) for the case $\alpha=0$, the most fluctuating collective-spin component transverse to the net magnetization keeps growing exponentially as 
$\max_\perp {{\rm Var}(J^\perp)} \approx \frac{N}{4} e^{2\lambda t}$, and up to longer times $t \sim \log N$ \cite{MunozArias2023}, reaching therefore \emph{Heisenberg scaling} of entanglement. This dynamics is observed for the entire range of fields $\Omega \in (0,{\cal J})$. Fig.~\ref{f.VarJyTaT}(b) shows the exponential growth of  ${\rm Var}(J^y)$ up to Heisenberg scaling in a time scaling as $\log N$; and the Heisenberg scaling of the maximum of ${\rm Var}(J^y)$ for the field range $\Omega \in (0,{\cal J})$ is shown in Fig.~\ref{f.VarJyTaT}(b). ${\rm Var}(J^y)$ corresponds to the quantum Fisher information for pure states, and its scaling as $N^2$ implies Heisenberg scaling of multipartite entanglement \cite{Pezze2018RMP}. Moreover it has a specific importance for an explicit metrological protocol based on the evolution of the parity under rotations \cite{FrerotR2024}, as we discuss further in the Conclusions (Sec.~\ref{s.conclusions}).
In fact Heisenberg scaling of the variance of all three components of the collective spins is observed, revealing that the collective-spin wavefunction  spreads broadly over the Bloch sphere.

\subsection{Dipolar interactions} 
The behavior of the collective spin fluctuations in the 2D dipolar case is instead more complex and intriguing. As shown in Fig.~\ref{f.VarJydip}, the fluctuations of {e.g.,} the component $J^y$ build up exponentially, and they clearly appear to reach a value consistent with Heisenberg scaling for a sufficiently small field ($\Omega/{\cal J} = 0.2$)---in complete analogy with what is seen in the $\alpha=0$ case. In particular, the maximum value of ${\rm Var}(J^y)$ lies above that reached by OAT-like dynamics ({i.e.,} for $\Omega=0$); and nonetheless it is reached in a time scaling only logarithmically with system size, so that TaT dynamics shows exponential acceleration towards Heisenberg scaling compared to OAT dynamics. The latter reaches indeed Heisenberg scaling in a much longer time, $t\sim \sqrt{N}$.

At variance with the $\alpha=0$ case, the onset of correlations has a peculiar spatial structure, which is rather remarkable. A microscopic insight into the spatial structure of the correlations for the $y$ spin components, $C^{yy}(d,t) = \langle S_{i}^y S_{i+d}^y\rangle(t)$, which sum up to ${\rm Var}(J^y)$, is obtained via our VMC, dTWA and RSW calculations---see Fig.~\ref{f.correlations}.  All three methods show that, in the dipolar system, correlations advance exponentially in time (i.e., nearly linearly for distances $d \ll L$ when plotted against $\exp(t)$). This represents the fastest propagation of correlations allowed by Lieb-Robinson bounds in systems with power-law interactions \cite{Tran2021,Tran2021_b,Lucas_review}. Hence dipolar TaT dynamics in 2D offers a paradigmatic  example of a simple time-independent Hamiltonian governing time-optimal ({i.e.,} exponential) propagation of quantum information for power-law interactions with $D < \alpha < 2D$. 

It is important to stress that this exponential buildup of long-range correlations has no relationship to thermalization dynamics, which is not at all observed in the time range we focused on. The thermal behavior of the TaT Hamiltonian is completely trivial:  at the temperature related to the energy of the initial state, the applied field, coupling to the order parameter of the system, very strongly polarizes the system, and correlations are short-ranged for all spin components. Using quantum Monte Carlo \cite{Sandvik2010AIPCP}, we calculate the thermodynamics expected at a temperature ($T/{\cal J} \approx 2$) corresponding to the energy of the initial coherent spin state for $\Omega/{\cal J} = 0.2$. 
The magnetization is expected to be $\langle J^x \rangle/N \approx -0.46$ completely at odds with anything to be seen in Fig.~\ref{f.squeezinga3}(d); and correlations are short-ranged, giving ${\rm Var}(J^y) \approx 5 N$. Therefore the dynamical onset of long-range correlations in this system is a genuine non-equilibrium effect. 

Nonetheless, when comparing the scaling of the maximum of ${\rm Var}(J^y)$ between the all-to-all case (Fig.~\ref{f.VarJyTaT}(c)) and the 2D dipolar case (Fig.~\ref{f.VarJydipscaling}) a clear difference emerges. In the dipolar case, Heisenberg scaling of the peak of ${\rm Var}(J^y)$ (as well as of the other collective-spin components) fails to persist after reaching a given system size---the smaller the larger the field. As seen in Fig.~\ref{f.VarJydipscaling}, dTWA predicts a well-defined crossover from Heisenberg scaling of the ${\rm Var}(J^y)$ peak to sub-Heisenberg scaling. This is a rather unconventional effect, which further underscores the non-equilibrium nature of the correlations associated with ${\rm Var}(J^y)$ dynamics. To quantitatively understand this crossover, we report its onset on the $(\Omega,  L)$ plane in Fig.~\ref{f.stability}: there we observe that the crossover line follows the same behavior as the line separating the regime of stable TaT dipolar dynamics from that of unstable dynamics due to the emergence of imaginary spin-wave frequencies. Therefore we conclude that the crossover in the scaling of the correlation peak relates to the onset of spin-wave instabilities, which appear to induce a strong deviation of the TaT dynamics in systems with power-law interactions compared to the case of all-to-all interactions. Spin-wave instabilities for large system sizes may also provide the mechanism by which TaT dynamics with power-law interactions eventually relaxes to a behavior aligning with its equilibrium thermodynamics of local observables \cite{dalessio_quantum_2016}.

\section{Conclusions} 
\label{s.conclusions}

In this work we have shown how twist-and-turn (TaT) dynamics in quantum spin Hamiltonians can lead to an exponential onset of entanglement in the form of squeezing at early times; and in the form of long-range correlations, transverse to the axis of initial polarization of the collective spin, at later times. Both observations are associated with scalable multipartite entanglement, which can be certified, as well as potentially exploited, using metrological criteria. Indeed squeezing is of immediate relevance to Ramsey interferometry \cite{Pezze2018RMP}.
Moreover, TaT dynamics has the special property of being constrained to a well-defined parity sector of the $J^x$ collective spin component. Under this condition,  the emergence of long-range correlations transverse to $J^x$ ({e.g.,} for the $S^y$ spin components) translates into an enhanced sensitivity of the above parity to rotations around the transverse axis ({e.g.,} the $J^y$ one), saturating the bound offered by the quantum Fisher information, and therefore reaching Heisenberg scaling in the TaT dynamics \cite{FrerotR2024}. 
The exponentially fast entanglement dynamics described above is observed for the quantum XY model both with all-to-all interactions, as well as for dipolar interactions in 2D. In the latter case, in spite of the model not being integrable, the observed entangling dynamics is at odds with thermalization, which would instead lead to an inversion of the collective spin without the development of long-range correlations. 
 
When compared with the paradigmatic one-axis-twisting dynamics, TaT dynamics offers an exponential acceleration of the onset of quantum correlations. The latter can be an asset for several experimental platforms with power-law interactions (neutral atoms and molecules, trapped ions, nitrogen vacancies, etc.)  in order to potentially generate entanglement at a faster rate than that related to the coupling of the system to its environment. 

\begin{acknowledgements}
This work is supported by PEPR-Q (QubitAF project). MK acknowledges the support from the Applied Quantum Computing Challenge Program at the
National Research Council of Canada. 
AC acknowledges support from the Canada First Research Excellence Fund and the Natural Sciences and Engineering Research Council of Canada.
All numerical simulations have been performed on the CBPsmn cluster at the ENS of Lyon. MK acknowledges useful discussions with Christopher Wyenberg and Kent Ueno. 
\end{acknowledgements}


\appendix

\section{From the twist-and-turn Hamiltonian to the squeezing Hamiltonian}
\label{a.bosonic}
In this section, we focus on the case of the all-to-all interactions ($\alpha=0$), and we connect the TaT Hamiltonian with the squeezing Hamiltonian of quantum optics via a linearized spin-boson transformation. 
The twist-and-turn Hamiltonian with $\alpha=0$ for $N$ spins with $S=1/2$ can be written in terms of the collective spin components as
\begin{equation}
H(\Omega) = \frac{(J^z)^2}{2I}  + \Omega J^x + {\rm const}
\end{equation}
where we have assumed that the collective spin length is maximal, $\bm J^2 = N/2(N/2+1)$, and therefore treated it as constant. Here the coupling constant reads $\frac{1}{2I} = \frac{\cal J}{N}$, to connect it with the notation of Eq.~\eqref{e.TaT}. 

In the following, we outline the spin-to-boson mapping, recall the general properties of the squeezing Hamiltonian in quantum optics, and show how the bosonic Hamiltonian---onto which the TaT Hamiltonian can be transformed---can, in general, be cast into the form of a squeezing Hamiltonian. This will allow us to deduce analytically the squeezing dynamics at short times, and to prove that the squeezed and anti-squeezed collective spin components evolve exponentially at a rate $\lambda = \sqrt{\Omega({\cal J}-\Omega)}$. 

\subsection{Spin-to-boson mapping}
\label{s.mapping}

The above Hamiltonian can be mapped onto a quadratic single-mode bosonic Hamiltonian, with operators $a, a^\dagger$, via the (linearized) Holstein-Primakoff transformation 
\begin{eqnarray}
J^x & = & \frac{N}{2} - a^\dagger a \nonumber \\
 J^y & \approx & \frac{\sqrt{N}}{2} \left (a + a^\dagger \right )  \nonumber \\  
 J^z & \approx  &  \frac{\sqrt{N}}{2i} \left ( a-a^\dagger \right ) 
\end{eqnarray}
to give
\begin{align}
& H(\Omega) \approx  H_b(\Omega, a, a^\dagger)=  \\ 
& \frac{\Omega N}{2} + \frac{N}{8I} - \frac{N}{8I} \left [ a^2 +( a^\dagger)^2 \right ] + \left ( \frac{N}{4I} - \Omega \right ) a^\dagger a + {\rm const}~. \nonumber
\end{align}
This mapping is valid as long the dynamics does not lead to a large deviation in the evolved state from the initial coherent spin state, such that $\langle J^x \rangle \lesssim N/2$, or $\langle a^\dagger a \rangle \ll N$. 

The above Hamiltonian reduces to the well-known \emph{squeezing Hamiltonian} in quantum optics, $H = - \frac{N}{8I} \left [ a^2 + (a^\dagger)^2 \right ] $,  at the optimal field 
\begin{equation}
\Omega_0 = \frac{N}{4I} = \frac{\cal J}{2}~. 
\end{equation}

The bosonic Hamiltonian (\emph{not} the spin Hamiltonian) has a fundamental \emph{symmetry}. If one reflects the field around the optimal value 
\begin{equation}
 \Omega  \rightarrow 2 \Omega_0 - \Omega
 \end{equation}
 and rotates the phase of the bosonic operators by $\pi/2$, namely $\tilde a = e^{i\pi/2} a$, then
 \begin{equation}
H_b(2\Omega_0 - \Omega, \tilde a, \tilde a^\dagger) = - H_b(\Omega, a, a^\dagger) + {\rm const.} 
\end{equation}
up to an irrelevant constant. The Hamiltonian is real valued in the Fock basis $|n\rangle$. When starting the evolution from a real-valued state in this basis---such as the vacuum, which corresponds to the coherent spin state $\otimes_i |\rightarrow_i\rangle$~--the subsequent evolution is time-reversal invariant, meaning that quantities that are also real-valued operators on the Fock basis have the same evolution under $H$ as under $-H$ \cite{Frerot2018PRL}. Hence the transformation $\Omega \rightarrow 2 \Omega_0 - \Omega$ and $\tilde a = e^{i\pi/2} a$ leaves the evolution unchanged, provided that one is comparing bosonic observables which are appropriately transformed (or are time-reversal symmetric).  

\subsection{Optimal field and bosonic squeezing}

At $\Omega = \Omega_0$ the bosonic Hamiltonian takes the squeezing form 
\begin{equation}
 H_b(\Omega_0, a, a^\dagger) = - \frac{N}{8I} \left [ a^2 +( a^\dagger)^2 \right ]  + {\rm const.}
\end{equation}
and the evolution operator is the squeezing operator 
\begin{align}
U(t) & = \exp[-iH_b(\Omega_0, a, a^\dagger) t] \nonumber \\ 
& = \exp\left ( \frac{q}{2}  (a^{\dagger})^2 - \frac{q^*}{2} a^2 \right ) = S(q)
\end{align}
with squeezing parameter $q  = i \frac{Nt}{4I} = \eta e^{i\phi}$  where $\eta = Nt/(4I) = {\cal J}t/2 $ and $\phi = \pi/2$~. 

As customarily done in quantum optics \cite{Agarwal-book}, we introduce the generalized mode quadrature 
\begin{equation}
X_\theta = \frac{e^{-i\theta} a+ e^{i\theta} a^\dagger}{\sqrt{2}} = \cos\theta ~X + \sin \theta~ P  
\end{equation}
where $X = X_0 = (a+a^\dagger)/\sqrt{2}$ and $P = X_{\pi/2} = (a - a^\dagger)/(i\sqrt{2})$. The application of the squeezing operator with $q = i \eta$ on the vacuum gives rise to a squeezed vacuum state, with a squeezed and anti-squeezed quadrature
\begin{equation}
{\rm Var}(X_{\pi/4}) = \frac{e^{2\eta}}{2} ~~~~~~~ {\rm Var}(X_{3\pi/4}) = \frac{e^{-2\eta}}{2}~.
\label{e.bosonic_squeezing}
\end{equation}
Moving back to spins, we obtain
\begin{equation}
J^y \approx \sqrt{\frac{N}{2}}~X ~~~~~~ J^z \approx \sqrt{\frac{N}{2}}~P,
\end{equation}
so that 
\begin{equation}
{\rm Var}\left ( \frac{J^y+J^z}{\sqrt{2}} \right )  \approx \frac{N}{4} e^{2\lambda t}~~~~~~~ {\rm Var}\left ( \frac{J^y-J^z}{\sqrt{2}} \right )  \approx \frac{N}{4} e^{-2\lambda t},
\end{equation}
where we have used the property $\eta = \lambda t = {\cal J}t/2$ at the optimal field $\Omega_0$.

On the other hand, the boson population evolves as
\begin{equation}
\langle a^\dagger a \rangle = \sinh(\eta)m
\end{equation}
yielding the spin squeezing parameter
\begin{equation}
\xi_R^2 = \frac{N{\rm Var}\left (\frac{J^y-J^z}{\sqrt{2}} \right )}{\langle J^x \rangle^2} \approx \frac{e^{-2\eta}}{\left[ 1-  \frac{2}{N}\sinh(\eta) \right ]^2}~.
\end{equation}
This result shows the \emph{exponential} onset of spin squeezing, and of exponential growth of the quantum Fisher information for the antisqueezed component, within the time frame of validity of the linearized spin-boson transformation---{i.e.,} as long as $\langle a^\dagger a \rangle = \sinh(\eta) \ll N$. 

\subsection{Squeezing away from the optimal field}
Away from the optimal field, the bosonic Hamiltonian has the form 
\begin{equation}
H_b = -\frac{\chi}{2} \left [ (a^\dagger)^2 + a^2 \right ] - \frac{\delta}{2} \left ( a^\dagger a + a a^\dagger \right )  
\end{equation}
where $\chi =  \frac{N}{4I}  = {\cal J}/2$ and $\delta = \Omega-\Omega_0 $. We now want to transform it into the squeezing Hamiltonian 
\begin{equation}
H_b = -\frac{\zeta}{2}   \left [ (b^\dagger)^2 + b^2 \right ] + {\rm const.} 
\end{equation}
via a \emph{squeezing} Bogolyubov transformation 
\begin{equation}
a = ub - v b^\dagger~~~~~~ b = u a + v a^\dagger~. 
\end{equation}
The transformation (with $|u|^2 - |v|^2=1$ to conserve the commutation relations) reads
\begin{equation}
 u = \sqrt{\frac{1}{2} \left (\frac{\chi}{\zeta} + 1 \right )}~~~~~~ v = {\rm sign}(\delta)~ \sqrt{\frac{1}{2} \left (\frac{\chi}{\zeta} - 1 \right )},
\end{equation}
where 
\begin{equation}
\zeta = \sqrt{\chi^2 - \delta^2} = \sqrt{\Omega ({\cal J} - \Omega)} = \lambda.
\end{equation}
This transformation leads to \emph{real} coefficients $u$ and $v$ under the condition
\begin{equation}
|\delta| \leq \chi  ~~\rightarrow ~~ |\Omega-\Omega_0| \leq \Omega_0 ~~~~~~ {\rm or} ~~~~~~~ \Omega \in [0, 2\Omega_0]~.
\label{e.condition}
\end{equation}
We shall restrict our attention to this interval, although the calculation can be generalized to values outside of it.  

The bosonic operators $b, b^\dagger$ are squeezed by $S(i\lambda t) = e^{-iH_b t}$ as \cite{Agarwal-book}
\begin{equation}
b(\lambda t) = S^\dagger b S = b \cosh(\lambda t) + i \sinh(\lambda t) b^\dagger,
\end{equation}
hence the transformation of the $a$ operator 
\begin{eqnarray}
a(\lambda) & = &  u~ b(\lambda t ) - v~ b^\dagger(\lambda t)  \\
& = & \left [ \cosh(\lambda t) + i \frac{\delta}{\lambda} \sinh( \lambda t) \right ]~a + i \frac{\chi}{\lambda} \sinh(\lambda t)~a^\dagger~.  \nonumber
\end{eqnarray}

After a lengthy but straightforward calculation, we obtain that the fluctuations of the generalized quadrature $X_\theta$ take the form 
\begin{eqnarray}
\langle X_{\theta}^2 \rangle & = &  \frac{1}{2} \frac{\delta}{\lambda^2} \left [ -\delta + \chi \cos (2\theta) \right ]  \nonumber \nonumber \\
& + & \frac{1}{4} \frac{\chi}{\lambda} \left [ \frac{\chi}{\lambda} + \sin(2\theta) - \frac{\delta}{\lambda} \cos(2\theta) \right ]  e^{2\lambda t} \nonumber \\
& + & \frac{1}{4} \frac{\chi}{\lambda} \left [ \frac{\chi}{\lambda} - \sin(2\theta) - \frac{\delta}{\lambda} \cos(2\theta) \right ]  e^{-2\lambda t},
\label{e.Xtheta}
\end{eqnarray}
which reproduces Eq.~\eqref{e.bosonic_squeezing} when taking $\delta = 0$ and $\theta = \{\pi/4, ~3\pi/4\}$. 

We observe the fundamental symmetry, highlighted in Sec.~\ref{s.mapping}
\begin{eqnarray}
\delta & \to &  -\delta \nonumber \\
\theta & \to & \theta + \pi/2  \nonumber \\
t &\to& -t ~.
\end{eqnarray} 

The boson number takes the form
\begin{equation}
\langle a^\dagger a \rangle = \left ( \frac{\chi}{\lambda} \right )^2 \sinh^2 (\lambda t),
\end{equation}
so that the spin squeezing parameter is
\begin{equation}
\xi_R^2 \approx \frac{2 \min_\theta \langle X_{\theta}^2 \rangle}{\left [ 1-\frac{2}{N} \left ( \frac{\chi}{\lambda} \right )^2 \sinh^2 (\lambda t) \right ]^2}~.
\label{e.xiboson}
\end{equation}

The above result (Eq.~\eqref{e.Xtheta}) shows that the anti-squeezed component of the collective spin grows exponentially as $e^{2\lambda t}$. At short time the state is of minimal uncertainty, so that the corresponding squeezed component decreases exponentially as   $e^{-2\lambda t}$. 

\subsection{Short-time behavior}
The short-time  expansion of  $\langle X_{\theta}^2 \rangle$ reads
\begin{equation}
\langle X_{\theta}^2 \rangle \approx \frac{1}{2} \left [ 1  + 2 \chi t \sin(2\theta) + 2 (\chi t)^2 - 2 \delta \chi t^2 \cos(2\theta) + O(t^3) \right ] 
\end{equation}
and that of the spin squeezing parameter reads 
\begin{equation}
\xi_R^2 \approx  \frac{ 1  - 2 \chi t  + O(t^2)}{\left ( 1 -  \frac{2\chi^2 t^2}{N} + O(t^4) \right )^2}
\end{equation}
namely the dependence on $\delta$ ({i.e.,} on the field) of the quadrature fluctuations and of the squeezing parameter only appears in \emph{quadratic terms} in time, $O(t^2)$. This implies that the time dependence of the squeezing parameter at short time is the same regardless of $\delta$---as long as it satisfies the condition of Eq.~\eqref{e.condition}. This result encompasses also the case $\delta = -\chi$ ({i.e.,} $\Omega =0$) corresponding to the one-axis-twisting dynamics. 

\subsection{Comparison between bosonic squeezing and spin squeezing}

\begin{figure}[ht!]
\begin{center}
\includegraphics[width=\columnwidth]{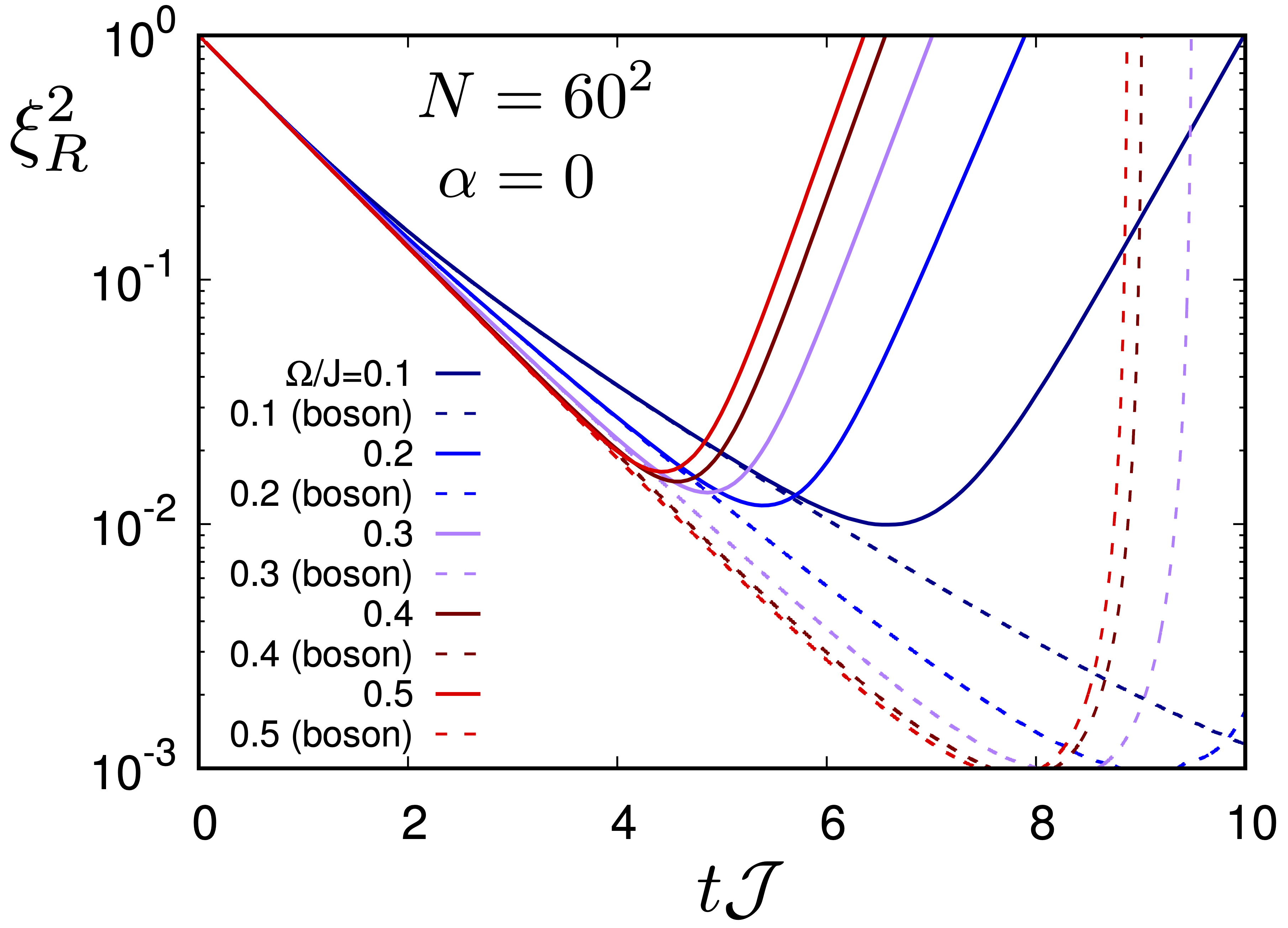}
\caption{Squeezing dynamics of the all-to-all TaT model for various fields (solid lines) compared with the predictions from the linearized bosonic approximation, Eq.~\eqref{e.xiboson} (dashed lines).}
\label{f.spinboson}
\end{center}
\end{figure}

Fig.~\ref{f.spinboson} presents an extended comparison between the predictions for squeezing from the bosonic model, Eq.~\eqref{e.xiboson}, and the full TaT dynamics for $\alpha=0$.  

We observe that the bosonic model predicts the correct squeezing dynamics down to the optimal squeezing, and in particular the $\Omega$-dependent exponential behavior which emerges after the $\Omega$-independent transient at short times (see previous subsection).  
Related to this crossover between short- and intermediate-time squeezing dynamics, there is an apparent change in the exponential decay dynamics of squeezing from a faster to a slower decay, observed at sufficiently small fields. This property, captured by the bosonic model for the case $\alpha=0$, is also seen in the case of the dipolar dynamics in 2D, as shown in Fig.~\ref{f.squeezinga3}(b).

In spite of its predictive power, the linearized bosonic model completely misses the value of the optimal squeezing observed for the spin model. We can therefore conclude that the observed optimal squeezing value stems fundamentally from the nonlinear nature of the spin dynamics on the Bloch sphere. 

\section{Rotor-spin-wave theory for the TaT Hamiltonian} 
\label{a.RSW}

Rotor-spin-wave (RSW) theory \cite{Roscildeetal2023} describes the dynamics of a quantum spin model in terms of two sets of variables: a rotor variable, which describes the dynamics projected onto the sector of Dicke states with maximal spin length ${\bm J}^2 = N/2(N/2+1)$; and spin-wave variables, which describe in a linearized fashion the dynamics leaking out of the Dicke sector. The theory relies on the fundamental (approximate) assumption that the two dynamics can be considered as decoupled, and it is valid under the condition that the population of the spin-wave modes remains small, allowing one to justifiably neglect the coupling between the spin waves and the rotor variable. 

Following the prescription of Refs.~\cite{Roscildeetal2023, Roscildeetal2023b}, we rewrite the TaT Hamiltonian ---Eq.~\eqref{e.TaT} --- as 
$H \approx H_{\rm R} + H_{\rm sw}$. Here
\begin{eqnarray}
H_{\rm R} & = &  - \frac{1}{2I} \left [  (K^x)^2+(K^y)^2 \right ] + \Omega K^x  \nonumber  \\
& = &  \frac{(K^z)^2}{2I} + \Omega K^x + {\rm const.} 
\end{eqnarray}
is the rotor Hamiltonian, where ${\bm K}$ is an angular momentum variable with ${\bm K}^2 = N/2(N/2+1)$ and 
\begin{equation}
\frac{1}{2I} = \frac{{\cal J} \gamma_0}{N-1} 
\end{equation}
is the effective inverse moment of inertia, with 
\begin{equation}
\gamma_{\bm k} = \sum_{\bm r \neq 0} \frac{e^{i \bm k \cdot \bm r}}{r^\alpha}~.  
\end{equation}
Moreover
\begin{equation}
H_{\rm sw} = \sum_{\bm k \neq 0} \omega_{\bm k} b_{\bm k}^\dagger b_{\bm k} + {\rm const.}
\end{equation}
is the spin-wave Hamiltonian, where $b_{\bm k}, b_{\bm k}^\dagger$ are Bogolyubov bosons diagonalizing the (linearized) bosonic Hamiltonian onto which the spin Hamiltonian is mapped via the Holstein-Primakoff transformatoin; and  $\omega_{\bm k} = \sqrt{A_{\bm k}^2 - B^2_{\bm k}}$ is the dispersion relation of the spin waves built around the initial coherent spin state, where \cite{Roscildeetal2023}:
\begin{equation}
A_{\bm k} =  {\cal J} ( \gamma_0 - \gamma_{\bm k}/2) - \Omega ~~~~~~ B_{\rm k} = - {\cal J} \gamma_{\bm k}/2~.
\end{equation}

The rotor Hamiltonian has the structure of a TaT Hamiltonian with $\alpha=0$, and therefore it drives squeezing dynamics with an exponential rate $\lambda_{\rm eff} = \sqrt{\Omega({\cal J}_{\rm eff} - \Omega)}$, where
\begin{equation}
{\cal J}_{\rm eff} = \frac{N}{2I} = \frac{N {\cal J} \gamma_0}{N-1} ~.
\end{equation}

When $\Omega = 0$, the spin-wave dispersion relation $\omega_{\bm k}$ has the characteristic structure of a gapless Goldstone mode, associated with the U(1) symmetry of the Hamiltonian. Upon application of the field, the frequency at zero wavevector becomes imaginary, namely $\omega_0 = \sqrt{-\Omega({\cal J} \gamma_0 - \Omega) } \approx i \lambda_{\rm eff}$ (for $N \gg 1$), reflecting the fact that the coherent spin state becomes an unstable fixed point of the dynamics. Yet, for sufficiently big sizes, also finite-wavevector spin-wave modes---which are the ones properly appearing in RSW theory---can acquire imaginary frequencies. On a $L\times L$ square lattice, this occurs first for the smallest wavevector ${\bm k} = (2\pi/L,0)$ when $A_{(2\pi/L,0)}^2 < B_{(2\pi/L,0)}^2$, namely
\begin{equation}
 \frac{\Omega}{\cal J} >  \gamma_0 - \gamma_{2\pi/L,0} ~ \sim L^{-2z}  ~~~~~ (L \gg 1)
\end{equation}
where $z = 1$ for $\alpha \geq D+2$, $z = (\alpha-D)/2$ for $D < \alpha < D+2$, and $z = 0$ for $\alpha \leq  D$ \cite{Frerot2017PRB}. 
For the dipolar interactions ($\alpha=3$) in $D=2$, which we have treated in this work, $z = 1/2$, so that the critical value of the field generating imaginary frequencies for finite-${\bm k}$ spin waves goes as $L^{-1}$---precisely the behavior shown in Fig.~\ref{f.stability}. 
Interestingly, for $\alpha < D$, the critical field for the onset of spin-wave instabilities does not scale with system size, meaning that there is an entire range of field values for which TaT dynamics is stable even in the thermodynamic limit, even though the system is not integrable; and even though the TaT dynamics producing squeezing and Heisenberg scaling of collective-spin fluctuations is incompatible with thermalization. 

On the other hand, the emergence of imaginary spin-wave frequencies leads to an exponential divergence of the populations of Holstein-Primakoff bosons, and to the breakdown of the RSW approach. Yet this breakdown occurs over time scales which are comparable with the inverse exponential divergence rate
\begin{eqnarray}
\lambda_{\bm k} & = &   \sqrt{\left | A_{\bm k}^2-B_{\bm k}^2 \right |}  \nonumber \\
& = & \sqrt{\left | \omega_{\bm k}^2(\Omega=0) - \Omega[ {\cal J}(2 \gamma_0 - \gamma_{\bm k}) - \Omega ] \right | } ~.
\end{eqnarray}
It is easy to observe that $\lambda_{\bm k} < \lambda_{\rm eff} = \lambda_{\bm k=0}$ for any finite ${\bm k}$, namely the rotor dynamics is systematically faster than the spin-wave dynamics, even in the presence of spin-wave instabilities. This mechanism may be responsible for the persistent scalability of squeezing observed in our simulations, even for system sizes and fields for which spin-wave instabilities (in the form of imaginary frequencies) are present. 

\begin{figure}[ht!]
\begin{center}
\includegraphics[width=0.75\columnwidth]{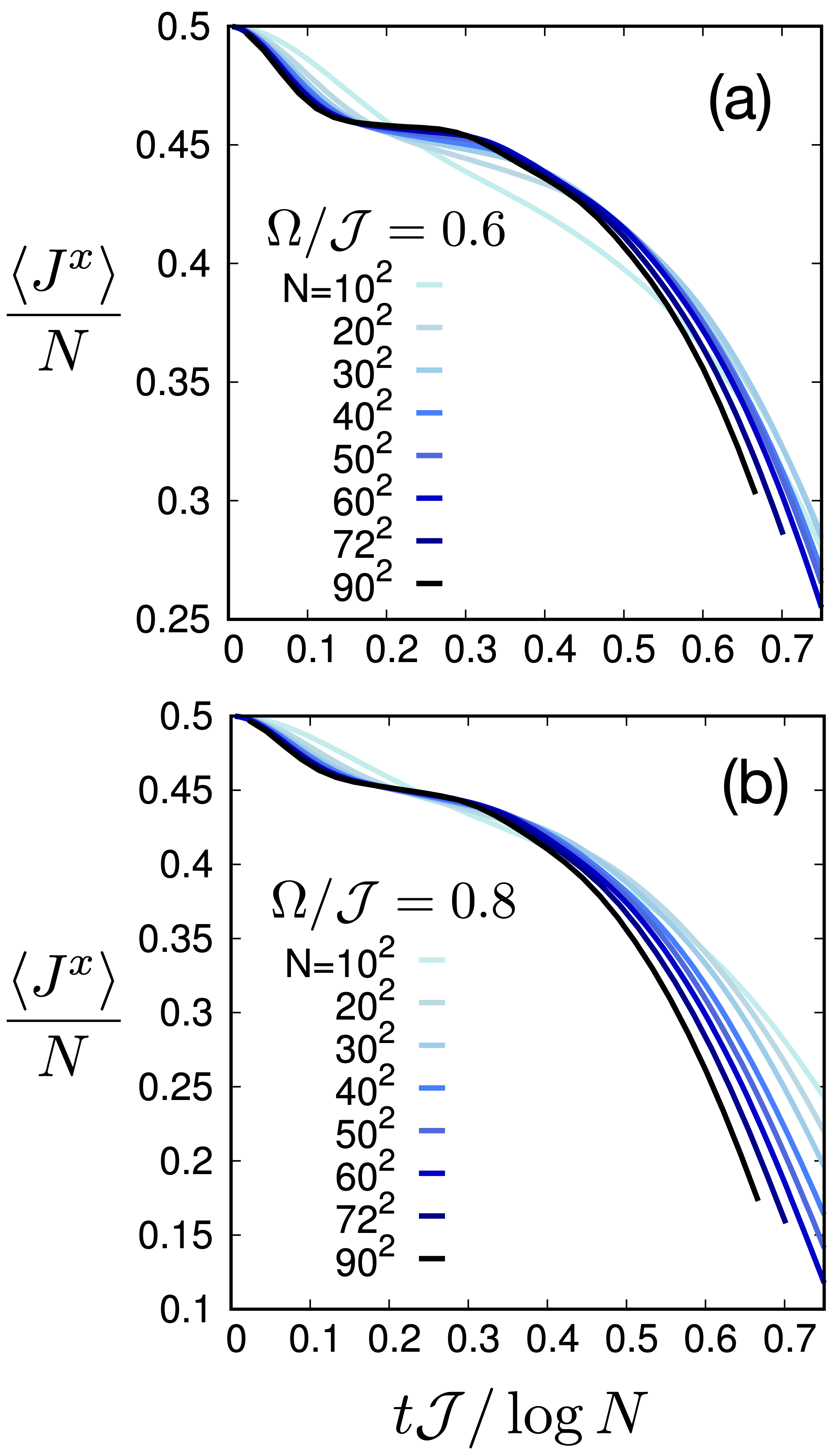}
\caption{Magnetization dynamics in the dipolar 2D system at large fields: (a) $\Omega = 0.6 \cal J$; (b) $\Omega = 0.8 \cal J$. All data are from dTWA calculations.}
\label{f.largeOm}
\end{center}
\end{figure}

\section{Magnetization dynamics for large system sizes and fields}
\label{a.Jx}

As seen in Sec.~\ref{s.dipsqueezing}, the dipolar system in 2D appears to exhibit a field-dependent scaling of optimal squeezing, at variance with the universal scaling $(\xi_R^2)_{\rm min} \sim N^{-1/2}$ seen in the $\alpha=0$ case. This could be interpreted as a consequence of the presence of spin-wave instabilities in the dipolar system, which are instead non-existent for $\alpha=0$ (in fact for any $\alpha<D$, as we argued in Sec.~\ref{s.SW} and in the previous section). Nonetheless this interpretation is incorrect. In the case of the dipolar interactions in 2D, and for all the system sizes studied in this work, the magnetization of the squeezed collective-spin component does not show any sign of instability over the time scale of squeezing dynamics. We illustrate this point by choosing the cases $\Omega/{\cal J} = 0.6$ and $\Omega/{\cal J} = 0.8$, which clearly possess spin-wave instabilities for the larger system size we investigated (see Sec.~\ref{s.SW}).

Within RSW theory, the magnetization should behave as 
$ \langle J^x \rangle = \langle K^x \rangle - \sum_{\bm k \neq 0} \langle a_{\bm k}^\dagger a_{\bm k} \rangle $, where $K^x$ is the $x$ component of the rotor variable introduced in App.~\ref{a.RSW}, following the all-to-all TaT dynamics; and $a_{\bm k}, a_{\bm k}^\dagger$  are Holstein-Primakoff bosons, related to the previously cited $b_{\bm k}, b_{\bm k}^\dagger$ bosons via a Bogolyubov transformation. A spin-wave instability would correspond to an exponential increase of the bosonic populations $\langle a_{\bm k}^\dagger a_{\bm k} \rangle$, corresponding to the wavevectors developing an imaginary frequency, and hence, to an exponential decrease of the magnetization. On the other hand, this behavior is not observed over the time scales of squeezing dynamics, as shown in Fig.~\ref{f.largeOm}. There we see that the magnetization remains macroscopic over timescales growing as $\log N$, i.e., the same behavior as in the case of smaller fields. 
This means that the change in the scaling of squeezing compared to the all-to-all TaT dynamics is fundamentally driven by the dynamics of the squeezed component of the collective spin transverse to the magnetization, which becomes field dependent. 

At the same time, it is interesting to observe that, after a size-independent plateau over times $\sim \log N$, the dynamics of the magnetization of the collective spin at larger fields starts showing significant differences with that observed at low fields (comparing Fig.~\ref{f.largeOm} with Fig.~\ref{f.squeezinga3}(a)). Indeed the depolarization dynamics becomes faster (in time units $\sim \log N$) the larger the system size,  likely a manifestation of the spin-wave instability. Investigating the nature of this dynamics at longer times, and the conditions under which it can lead to thermalization, will be the subject of future work. 

\bibliography{squeezing.bib}

\begin{thebibliography}{35}
\expandafter\ifx\csname natexlab\endcsname\relax\def\natexlab#1{#1}\fi
\expandafter\ifx\csname bibnamefont\endcsname\relax
  \def\bibnamefont#1{#1}\fi
\expandafter\ifx\csname bibfnamefont\endcsname\relax
  \def\bibfnamefont#1{#1}\fi
\expandafter\ifx\csname citenamefont\endcsname\relax
  \def\citenamefont#1{#1}\fi
\expandafter\ifx\csname url\endcsname\relax
  \def\url#1{\texttt{#1}}\fi
\expandafter\ifx\csname urlprefix\endcsname\relax\def\urlprefix{URL }\fi
\providecommand{\bibinfo}[2]{#2}
\providecommand{\eprint}[2][]{\url{#2}}

\bibitem[{\citenamefont{D'Alessio et~al.}(2016)\citenamefont{D'Alessio, Kafri,
  Polkovnikov, and Rigol}}]{dalessio_quantum_2016}
\bibinfo{author}{\bibfnamefont{L.}~\bibnamefont{D'Alessio}},
  \bibinfo{author}{\bibfnamefont{Y.}~\bibnamefont{Kafri}},
  \bibinfo{author}{\bibfnamefont{A.}~\bibnamefont{Polkovnikov}},
  \bibnamefont{and} \bibinfo{author}{\bibfnamefont{M.}~\bibnamefont{Rigol}},
  \bibinfo{journal}{Advances in Physics} \textbf{\bibinfo{volume}{65}},
  \bibinfo{pages}{239} (\bibinfo{year}{2016}), ISSN \bibinfo{issn}{0001-8732},
  \urlprefix\url{https://doi.org/10.1080/00018732.2016.1198134}.

\bibitem[{\citenamefont{Kaufman et~al.}(2016)\citenamefont{Kaufman, Tai, Lukin,
  Rispoli, Schittko, Preiss, and Greiner}}]{Kaufmanetal2016}
\bibinfo{author}{\bibfnamefont{A.~M.} \bibnamefont{Kaufman}},
  \bibinfo{author}{\bibfnamefont{M.~E.} \bibnamefont{Tai}},
  \bibinfo{author}{\bibfnamefont{A.}~\bibnamefont{Lukin}},
  \bibinfo{author}{\bibfnamefont{M.}~\bibnamefont{Rispoli}},
  \bibinfo{author}{\bibfnamefont{R.}~\bibnamefont{Schittko}},
  \bibinfo{author}{\bibfnamefont{P.~M.} \bibnamefont{Preiss}},
  \bibnamefont{and} \bibinfo{author}{\bibfnamefont{M.}~\bibnamefont{Greiner}},
  \bibinfo{journal}{Science} \textbf{\bibinfo{volume}{353}},
  \bibinfo{pages}{794} (\bibinfo{year}{2016}), ISSN \bibinfo{issn}{0036-8075},
  \urlprefix\url{http://science.sciencemag.org/content/353/6301/794}.

\bibitem[{\citenamefont{Abanin et~al.}(2019)\citenamefont{Abanin, Altman,
  Bloch, and Serbyn}}]{Abaninetal2019}
\bibinfo{author}{\bibfnamefont{D.~A.} \bibnamefont{Abanin}},
  \bibinfo{author}{\bibfnamefont{E.}~\bibnamefont{Altman}},
  \bibinfo{author}{\bibfnamefont{I.}~\bibnamefont{Bloch}}, \bibnamefont{and}
  \bibinfo{author}{\bibfnamefont{M.}~\bibnamefont{Serbyn}},
  \bibinfo{journal}{Rev. Mod. Phys.} \textbf{\bibinfo{volume}{91}},
  \bibinfo{pages}{021001} (\bibinfo{year}{2019}),
  \urlprefix\url{https://link.aps.org/doi/10.1103/RevModPhys.91.021001}.

\bibitem[{\citenamefont{G{\"a}rttner et~al.}(2017)\citenamefont{G{\"a}rttner,
  Bohnet, Safavi-Naini, Wall, Bollinger, and Rey}}]{Garttner2017}
\bibinfo{author}{\bibfnamefont{M.}~\bibnamefont{G{\"a}rttner}},
  \bibinfo{author}{\bibfnamefont{J.~G.} \bibnamefont{Bohnet}},
  \bibinfo{author}{\bibfnamefont{A.}~\bibnamefont{Safavi-Naini}},
  \bibinfo{author}{\bibfnamefont{M.~L.} \bibnamefont{Wall}},
  \bibinfo{author}{\bibfnamefont{J.~J.} \bibnamefont{Bollinger}},
  \bibnamefont{and} \bibinfo{author}{\bibfnamefont{A.~M.} \bibnamefont{Rey}},
  \bibinfo{journal}{Nature Physics} \textbf{\bibinfo{volume}{13}},
  \bibinfo{pages}{781} (\bibinfo{year}{2017}), ISSN \bibinfo{issn}{1745-2481},
  \urlprefix\url{https://doi.org/10.1038/nphys4119}.

\bibitem[{\citenamefont{Li et~al.}(2023)\citenamefont{Li, null, Colombo, Shu,
  Velez, Pilatowsky-Cameo, Schmied, Choi, Lukin, Pedrozo-Pe{\~{n}}afiel
  et~al.}}]{Lietal2023}
\bibinfo{author}{\bibfnamefont{Z.}~\bibnamefont{Li}},
  \bibinfo{author}{\bibnamefont{null}},
  \bibinfo{author}{\bibfnamefont{S.}~\bibnamefont{Colombo}},
  \bibinfo{author}{\bibfnamefont{C.}~\bibnamefont{Shu}},
  \bibinfo{author}{\bibfnamefont{G.}~\bibnamefont{Velez}},
  \bibinfo{author}{\bibfnamefont{S.}~\bibnamefont{Pilatowsky-Cameo}},
  \bibinfo{author}{\bibfnamefont{R.}~\bibnamefont{Schmied}},
  \bibinfo{author}{\bibfnamefont{S.}~\bibnamefont{Choi}},
  \bibinfo{author}{\bibfnamefont{M.}~\bibnamefont{Lukin}},
  \bibinfo{author}{\bibfnamefont{E.}~\bibnamefont{Pedrozo-Pe{\~{n}}afiel}},
  \bibnamefont{et~al.}, \bibinfo{journal}{Science}
  \textbf{\bibinfo{volume}{380}}, \bibinfo{pages}{1381} (\bibinfo{year}{2023}),
  \eprint{https://www.science.org/doi/pdf/10.1126/science.adg9500},
  \urlprefix\url{https://www.science.org/doi/abs/10.1126/science.adg9500}.

\bibitem[{\citenamefont{Xu and Swingle}(2024)}]{XuS2024}
\bibinfo{author}{\bibfnamefont{S.}~\bibnamefont{Xu}} \bibnamefont{and}
  \bibinfo{author}{\bibfnamefont{B.}~\bibnamefont{Swingle}},
  \bibinfo{journal}{PRX Quantum} \textbf{\bibinfo{volume}{5}},
  \bibinfo{pages}{010201} (\bibinfo{year}{2024}),
  \urlprefix\url{https://link.aps.org/doi/10.1103/PRXQuantum.5.010201}.

\bibitem[{\citenamefont{Chen et~al.}(2023)\citenamefont{Chen, Lucas, and
  Yin}}]{Lucas_review}
\bibinfo{author}{\bibfnamefont{C.-F.~A.} \bibnamefont{Chen}},
  \bibinfo{author}{\bibfnamefont{A.}~\bibnamefont{Lucas}}, \bibnamefont{and}
  \bibinfo{author}{\bibfnamefont{C.}~\bibnamefont{Yin}},
  \bibinfo{journal}{Reports on Progress in Physics}
  \textbf{\bibinfo{volume}{86}}, \bibinfo{pages}{116001}
  (\bibinfo{year}{2023}),
  \urlprefix\url{https://dx.doi.org/10.1088/1361-6633/acfaae}.

\bibitem[{\citenamefont{Cheneau et~al.}(2012)\citenamefont{Cheneau, Barmettler,
  Poletti, Endres, Schau{\ss}, Fukuhara, Gross, Bloch, Kollath, and
  Kuhr}}]{Cheneauetal2012}
\bibinfo{author}{\bibfnamefont{M.}~\bibnamefont{Cheneau}},
  \bibinfo{author}{\bibfnamefont{P.}~\bibnamefont{Barmettler}},
  \bibinfo{author}{\bibfnamefont{D.}~\bibnamefont{Poletti}},
  \bibinfo{author}{\bibfnamefont{M.}~\bibnamefont{Endres}},
  \bibinfo{author}{\bibfnamefont{P.}~\bibnamefont{Schau{\ss}}},
  \bibinfo{author}{\bibfnamefont{T.}~\bibnamefont{Fukuhara}},
  \bibinfo{author}{\bibfnamefont{C.}~\bibnamefont{Gross}},
  \bibinfo{author}{\bibfnamefont{I.}~\bibnamefont{Bloch}},
  \bibinfo{author}{\bibfnamefont{C.}~\bibnamefont{Kollath}}, \bibnamefont{and}
  \bibinfo{author}{\bibfnamefont{S.}~\bibnamefont{Kuhr}},
  \bibinfo{journal}{Nature} \textbf{\bibinfo{volume}{481}},
  \bibinfo{pages}{484} (\bibinfo{year}{2012}), ISSN \bibinfo{issn}{1476-4687},
  \urlprefix\url{https://doi.org/10.1038/nature10748}.

\bibitem[{\citenamefont{Carleo et~al.}(2014)\citenamefont{Carleo, Becca,
  Sanchez-Palencia, Sorella, and Fabrizio}}]{Carleoetal2014}
\bibinfo{author}{\bibfnamefont{G.}~\bibnamefont{Carleo}},
  \bibinfo{author}{\bibfnamefont{F.}~\bibnamefont{Becca}},
  \bibinfo{author}{\bibfnamefont{L.}~\bibnamefont{Sanchez-Palencia}},
  \bibinfo{author}{\bibfnamefont{S.}~\bibnamefont{Sorella}}, \bibnamefont{and}
  \bibinfo{author}{\bibfnamefont{M.}~\bibnamefont{Fabrizio}},
  \bibinfo{journal}{Phys. Rev. A} \textbf{\bibinfo{volume}{89}},
  \bibinfo{pages}{031602} (\bibinfo{year}{2014}),
  \urlprefix\url{https://link.aps.org/doi/10.1103/PhysRevA.89.031602}.

\bibitem[{\citenamefont{Fr\'erot et~al.}(2018)\citenamefont{Fr\'erot, Naldesi,
  and Roscilde}}]{Frerot2018PRL}
\bibinfo{author}{\bibfnamefont{I.}~\bibnamefont{Fr\'erot}},
  \bibinfo{author}{\bibfnamefont{P.}~\bibnamefont{Naldesi}}, \bibnamefont{and}
  \bibinfo{author}{\bibfnamefont{T.}~\bibnamefont{Roscilde}},
  \bibinfo{journal}{Phys. Rev. Lett.} \textbf{\bibinfo{volume}{120}},
  \bibinfo{pages}{050401} (\bibinfo{year}{2018}),
  \urlprefix\url{https://link.aps.org/doi/10.1103/PhysRevLett.120.050401}.

\bibitem[{\citenamefont{Cevolani et~al.}(2018)\citenamefont{Cevolani, Despres,
  Carleo, Tagliacozzo, and Sanchez-Palencia}}]{Cevolanietal2019}
\bibinfo{author}{\bibfnamefont{L.}~\bibnamefont{Cevolani}},
  \bibinfo{author}{\bibfnamefont{J.}~\bibnamefont{Despres}},
  \bibinfo{author}{\bibfnamefont{G.}~\bibnamefont{Carleo}},
  \bibinfo{author}{\bibfnamefont{L.}~\bibnamefont{Tagliacozzo}},
  \bibnamefont{and}
  \bibinfo{author}{\bibfnamefont{L.}~\bibnamefont{Sanchez-Palencia}},
  \bibinfo{journal}{Phys. Rev. B} \textbf{\bibinfo{volume}{98}},
  \bibinfo{pages}{024302} (\bibinfo{year}{2018}),
  \urlprefix\url{https://link.aps.org/doi/10.1103/PhysRevB.98.024302}.

\bibitem[{\citenamefont{Hastings and Koma}(2006)}]{HastingsK2006}
\bibinfo{author}{\bibfnamefont{M.~B.} \bibnamefont{Hastings}} \bibnamefont{and}
  \bibinfo{author}{\bibfnamefont{T.}~\bibnamefont{Koma}},
  \bibinfo{journal}{Communications in Mathematical Physics}
  \textbf{\bibinfo{volume}{265}}, \bibinfo{pages}{781} (\bibinfo{year}{2006}),
  ISSN \bibinfo{issn}{1432-0916},
  \urlprefix\url{https://doi.org/10.1007/s00220-006-0030-4}.

\bibitem[{\citenamefont{Tran et~al.}(2021{\natexlab{a}})\citenamefont{Tran,
  Guo, Baldwin, Ehrenberg, Gorshkov, and Lucas}}]{Tran2021_b}
\bibinfo{author}{\bibfnamefont{M.~C.} \bibnamefont{Tran}},
  \bibinfo{author}{\bibfnamefont{A.~Y.} \bibnamefont{Guo}},
  \bibinfo{author}{\bibfnamefont{C.~L.} \bibnamefont{Baldwin}},
  \bibinfo{author}{\bibfnamefont{A.}~\bibnamefont{Ehrenberg}},
  \bibinfo{author}{\bibfnamefont{A.~V.} \bibnamefont{Gorshkov}},
  \bibnamefont{and} \bibinfo{author}{\bibfnamefont{A.}~\bibnamefont{Lucas}},
  \bibinfo{journal}{Phys. Rev. Lett.} \textbf{\bibinfo{volume}{127}},
  \bibinfo{pages}{160401} (\bibinfo{year}{2021}{\natexlab{a}}),
  \urlprefix\url{https://link.aps.org/doi/10.1103/PhysRevLett.127.160401}.

\bibitem[{\citenamefont{Tran et~al.}(2021{\natexlab{b}})\citenamefont{Tran,
  Guo, Deshpande, Lucas, and Gorshkov}}]{Tran2021}
\bibinfo{author}{\bibfnamefont{M.~C.} \bibnamefont{Tran}},
  \bibinfo{author}{\bibfnamefont{A.~Y.} \bibnamefont{Guo}},
  \bibinfo{author}{\bibfnamefont{A.}~\bibnamefont{Deshpande}},
  \bibinfo{author}{\bibfnamefont{A.}~\bibnamefont{Lucas}}, \bibnamefont{and}
  \bibinfo{author}{\bibfnamefont{A.~V.} \bibnamefont{Gorshkov}},
  \bibinfo{journal}{Phys. Rev. X} \textbf{\bibinfo{volume}{11}},
  \bibinfo{pages}{031016} (\bibinfo{year}{2021}{\natexlab{b}}),
  \urlprefix\url{https://link.aps.org/doi/10.1103/PhysRevX.11.031016}.

\bibitem[{\citenamefont{Polkovnikov et~al.}(2011)\citenamefont{Polkovnikov,
  Sengupta, Silva, and Vengalattore}}]{Polkovnikov2011}
\bibinfo{author}{\bibfnamefont{A.}~\bibnamefont{Polkovnikov}},
  \bibinfo{author}{\bibfnamefont{K.}~\bibnamefont{Sengupta}},
  \bibinfo{author}{\bibfnamefont{A.}~\bibnamefont{Silva}}, \bibnamefont{and}
  \bibinfo{author}{\bibfnamefont{M.}~\bibnamefont{Vengalattore}},
  \bibinfo{journal}{Rev. Mod. Phys.} \textbf{\bibinfo{volume}{83}},
  \bibinfo{pages}{863} (\bibinfo{year}{2011}),
  \urlprefix\url{https://link.aps.org/doi/10.1103/RevModPhys.83.863}.

\bibitem[{\citenamefont{Micheli et~al.}(2003)\citenamefont{Micheli, Jaksch,
  Cirac, and Zoller}}]{Michelietal2003}
\bibinfo{author}{\bibfnamefont{A.}~\bibnamefont{Micheli}},
  \bibinfo{author}{\bibfnamefont{D.}~\bibnamefont{Jaksch}},
  \bibinfo{author}{\bibfnamefont{J.~I.} \bibnamefont{Cirac}}, \bibnamefont{and}
  \bibinfo{author}{\bibfnamefont{P.}~\bibnamefont{Zoller}},
  \bibinfo{journal}{Phys. Rev. A} \textbf{\bibinfo{volume}{67}},
  \bibinfo{pages}{013607} (\bibinfo{year}{2003}),
  \urlprefix\url{https://link.aps.org/doi/10.1103/PhysRevA.67.013607}.

\bibitem[{\citenamefont{Muessel et~al.}(2015)\citenamefont{Muessel, Strobel,
  Linnemann, Zibold, Juli\'a-D\'{\i}az, and Oberthaler}}]{Muesseletal2015}
\bibinfo{author}{\bibfnamefont{W.}~\bibnamefont{Muessel}},
  \bibinfo{author}{\bibfnamefont{H.}~\bibnamefont{Strobel}},
  \bibinfo{author}{\bibfnamefont{D.}~\bibnamefont{Linnemann}},
  \bibinfo{author}{\bibfnamefont{T.}~\bibnamefont{Zibold}},
  \bibinfo{author}{\bibfnamefont{B.}~\bibnamefont{Juli\'a-D\'{\i}az}},
  \bibnamefont{and} \bibinfo{author}{\bibfnamefont{M.~K.}
  \bibnamefont{Oberthaler}}, \bibinfo{journal}{Phys. Rev. A}
  \textbf{\bibinfo{volume}{92}}, \bibinfo{pages}{023603}
  (\bibinfo{year}{2015}),
  \urlprefix\url{https://link.aps.org/doi/10.1103/PhysRevA.92.023603}.

\bibitem[{\citenamefont{Sorelli et~al.}(2019)\citenamefont{Sorelli, Gessner,
  Smerzi, and Pezz\`e}}]{Sorellietal2019}
\bibinfo{author}{\bibfnamefont{G.}~\bibnamefont{Sorelli}},
  \bibinfo{author}{\bibfnamefont{M.}~\bibnamefont{Gessner}},
  \bibinfo{author}{\bibfnamefont{A.}~\bibnamefont{Smerzi}}, \bibnamefont{and}
  \bibinfo{author}{\bibfnamefont{L.}~\bibnamefont{Pezz\`e}},
  \bibinfo{journal}{Phys. Rev. A} \textbf{\bibinfo{volume}{99}},
  \bibinfo{pages}{022329} (\bibinfo{year}{2019}),
  \urlprefix\url{https://link.aps.org/doi/10.1103/PhysRevA.99.022329}.

\bibitem[{\citenamefont{Mu{\~{n}}oz-Arias
  et~al.}(2023)\citenamefont{Mu{\~{n}}oz-Arias, Deutsch, and
  Poggi}}]{MunozArias2023}
\bibinfo{author}{\bibfnamefont{M.~H.} \bibnamefont{Mu{\~{n}}oz-Arias}},
  \bibinfo{author}{\bibfnamefont{I.~H.} \bibnamefont{Deutsch}},
  \bibnamefont{and} \bibinfo{author}{\bibfnamefont{P.~M.} \bibnamefont{Poggi}},
  \bibinfo{journal}{PRX Quantum} \textbf{\bibinfo{volume}{4}},
  \bibinfo{pages}{020314} (\bibinfo{year}{2023}),
  \urlprefix\url{https://link.aps.org/doi/10.1103/PRXQuantum.4.020314}.

\bibitem[{\citenamefont{Strobel et~al.}(2014)\citenamefont{Strobel, Muessel,
  Linnemann, Zibold, Hume, Pezz\`e, Smerzi, and Oberthaler}}]{Strobeletal2014}
\bibinfo{author}{\bibfnamefont{H.}~\bibnamefont{Strobel}},
  \bibinfo{author}{\bibfnamefont{W.}~\bibnamefont{Muessel}},
  \bibinfo{author}{\bibfnamefont{D.}~\bibnamefont{Linnemann}},
  \bibinfo{author}{\bibfnamefont{T.}~\bibnamefont{Zibold}},
  \bibinfo{author}{\bibfnamefont{D.~B.} \bibnamefont{Hume}},
  \bibinfo{author}{\bibfnamefont{L.}~\bibnamefont{Pezz\`e}},
  \bibinfo{author}{\bibfnamefont{A.}~\bibnamefont{Smerzi}}, \bibnamefont{and}
  \bibinfo{author}{\bibfnamefont{M.~K.} \bibnamefont{Oberthaler}},
  \bibinfo{journal}{Science} \textbf{\bibinfo{volume}{345}},
  \bibinfo{pages}{424} (\bibinfo{year}{2014}),
  \eprint{https://www.science.org/doi/pdf/10.1126/science.1250147},
  \urlprefix\url{https://www.science.org/doi/abs/10.1126/science.1250147}.

\bibitem[{\citenamefont{Browaeys and Lahaye}(2020)}]{BrowaeysL2020}
\bibinfo{author}{\bibfnamefont{A.}~\bibnamefont{Browaeys}} \bibnamefont{and}
  \bibinfo{author}{\bibfnamefont{T.}~\bibnamefont{Lahaye}},
  \bibinfo{journal}{Nat. Phys.} \textbf{\bibinfo{volume}{16}},
  \bibinfo{pages}{132} (\bibinfo{year}{2020}),
  \urlprefix\url{https://doi.org/10.1038/s41567-019-0733-z}.

\bibitem[{\citenamefont{Chomaz et~al.}(2022)\citenamefont{Chomaz,
  Ferrier-Barbut, Ferlaino, Laburthe-Tolra, Lev, and Pfau}}]{Chomazetal2022}
\bibinfo{author}{\bibfnamefont{L.}~\bibnamefont{Chomaz}},
  \bibinfo{author}{\bibfnamefont{I.}~\bibnamefont{Ferrier-Barbut}},
  \bibinfo{author}{\bibfnamefont{F.}~\bibnamefont{Ferlaino}},
  \bibinfo{author}{\bibfnamefont{B.}~\bibnamefont{Laburthe-Tolra}},
  \bibinfo{author}{\bibfnamefont{B.~L.} \bibnamefont{Lev}}, \bibnamefont{and}
  \bibinfo{author}{\bibfnamefont{T.}~\bibnamefont{Pfau}},
  \bibinfo{journal}{Reports on Progress in Physics}
  \textbf{\bibinfo{volume}{86}}, \bibinfo{pages}{026401}
  (\bibinfo{year}{2022}),
  \urlprefix\url{https://dx.doi.org/10.1088/1361-6633/aca814}.

\bibitem[{\citenamefont{Cornish et~al.}(2024)\citenamefont{Cornish, Tarbutt,
  and Hazzard}}]{Cornish2024}
\bibinfo{author}{\bibfnamefont{S.~L.} \bibnamefont{Cornish}},
  \bibinfo{author}{\bibfnamefont{M.~R.} \bibnamefont{Tarbutt}},
  \bibnamefont{and} \bibinfo{author}{\bibfnamefont{K.~R.~A.}
  \bibnamefont{Hazzard}}, \bibinfo{journal}{Nature Physics}
  \textbf{\bibinfo{volume}{20}}, \bibinfo{pages}{730} (\bibinfo{year}{2024}),
  ISSN \bibinfo{issn}{1745-2481},
  \urlprefix\url{https://doi.org/10.1038/s41567-024-02453-9}.

\bibitem[{\citenamefont{Fr\'erot and Roscilde}(2024)}]{FrerotR2024}
\bibinfo{author}{\bibfnamefont{I.}~\bibnamefont{Fr\'erot}} \bibnamefont{and}
  \bibinfo{author}{\bibfnamefont{T.}~\bibnamefont{Roscilde}},
  \bibinfo{journal}{Phys. Rev. Lett.} \textbf{\bibinfo{volume}{133}},
  \bibinfo{pages}{260402} (\bibinfo{year}{2024}),
  \urlprefix\url{https://link.aps.org/doi/10.1103/PhysRevLett.133.260402}.

\bibitem[{\citenamefont{Pezz\`e et~al.}(2018)\citenamefont{Pezz\`e, Smerzi,
  Oberthaler, Schmied, and Treutlein}}]{Pezze2018RMP}
\bibinfo{author}{\bibfnamefont{L.}~\bibnamefont{Pezz\`e}},
  \bibinfo{author}{\bibfnamefont{A.}~\bibnamefont{Smerzi}},
  \bibinfo{author}{\bibfnamefont{M.~K.} \bibnamefont{Oberthaler}},
  \bibinfo{author}{\bibfnamefont{R.}~\bibnamefont{Schmied}}, \bibnamefont{and}
  \bibinfo{author}{\bibfnamefont{P.}~\bibnamefont{Treutlein}},
  \bibinfo{journal}{Rev. Mod. Phys.} \textbf{\bibinfo{volume}{90}},
  \bibinfo{pages}{035005} (\bibinfo{year}{2018}),
  \urlprefix\url{https://link.aps.org/doi/10.1103/RevModPhys.90.035005}.

\bibitem[{\citenamefont{Schachenmayer et~al.}(2015)\citenamefont{Schachenmayer,
  Pikovski, and Rey}}]{Schachenmayer2015PRX}
\bibinfo{author}{\bibfnamefont{J.}~\bibnamefont{Schachenmayer}},
  \bibinfo{author}{\bibfnamefont{A.}~\bibnamefont{Pikovski}}, \bibnamefont{and}
  \bibinfo{author}{\bibfnamefont{A.~M.} \bibnamefont{Rey}},
  \bibinfo{journal}{Phys. Rev. X} \textbf{\bibinfo{volume}{5}},
  \bibinfo{pages}{011022} (\bibinfo{year}{2015}),
  \urlprefix\url{https://link.aps.org/doi/10.1103/PhysRevX.5.011022}.

\bibitem[{\citenamefont{Thibaut et~al.}(2019)\citenamefont{Thibaut, Roscilde,
  and Mezzacapo}}]{PRB2019}
\bibinfo{author}{\bibfnamefont{J.}~\bibnamefont{Thibaut}},
  \bibinfo{author}{\bibfnamefont{T.}~\bibnamefont{Roscilde}}, \bibnamefont{and}
  \bibinfo{author}{\bibfnamefont{F.}~\bibnamefont{Mezzacapo}},
  \bibinfo{journal}{Phys. Rev. B} \textbf{\bibinfo{volume}{100}},
  \bibinfo{pages}{155148} (\bibinfo{year}{2019}),
  \urlprefix\url{https://link.aps.org/doi/10.1103/PhysRevB.100.155148}.

\bibitem[{\citenamefont{Comparin et~al.}(2022)\citenamefont{Comparin,
  Mezzacapo, and Roscilde}}]{Comparin2022PRA}
\bibinfo{author}{\bibfnamefont{T.}~\bibnamefont{Comparin}},
  \bibinfo{author}{\bibfnamefont{F.}~\bibnamefont{Mezzacapo}},
  \bibnamefont{and} \bibinfo{author}{\bibfnamefont{T.}~\bibnamefont{Roscilde}},
  \bibinfo{journal}{Phys. Rev. A} \textbf{\bibinfo{volume}{105}},
  \bibinfo{pages}{022625} (\bibinfo{year}{2022}),
  \urlprefix\url{https://link.aps.org/doi/10.1103/PhysRevA.105.022625}.

\bibitem[{\citenamefont{Caleca et~al.}(2024)\citenamefont{Caleca, Bocini,
  Mezzacapo, and Roscilde}}]{Calecaetal2024}
\bibinfo{author}{\bibfnamefont{F.}~\bibnamefont{Caleca}},
  \bibinfo{author}{\bibfnamefont{S.}~\bibnamefont{Bocini}},
  \bibinfo{author}{\bibfnamefont{F.}~\bibnamefont{Mezzacapo}},
  \bibnamefont{and} \bibinfo{author}{\bibfnamefont{T.}~\bibnamefont{Roscilde}},
  \emph{\bibinfo{title}{Giant number-parity effect leading to spontaneous
  symmetry breaking in finite-size quantum spin models}}
  (\bibinfo{year}{2024}), \eprint{2412.15493},
  \urlprefix\url{https://arxiv.org/abs/2412.15493}.

\bibitem[{\citenamefont{Roscilde
  et~al.}(2023{\natexlab{a}})\citenamefont{Roscilde, Comparin, and
  Mezzacapo}}]{Roscildeetal2023}
\bibinfo{author}{\bibfnamefont{T.}~\bibnamefont{Roscilde}},
  \bibinfo{author}{\bibfnamefont{T.}~\bibnamefont{Comparin}}, \bibnamefont{and}
  \bibinfo{author}{\bibfnamefont{F.}~\bibnamefont{Mezzacapo}},
  \bibinfo{journal}{Phys. Rev. Lett.} \textbf{\bibinfo{volume}{131}},
  \bibinfo{pages}{160403} (\bibinfo{year}{2023}{\natexlab{a}}),
  \urlprefix\url{https://link.aps.org/doi/10.1103/PhysRevLett.131.160403}.

\bibitem[{\citenamefont{Fr\'erot et~al.}(2017)\citenamefont{Fr\'erot, Naldesi,
  and Roscilde}}]{Frerot2017PRB}
\bibinfo{author}{\bibfnamefont{I.}~\bibnamefont{Fr\'erot}},
  \bibinfo{author}{\bibfnamefont{P.}~\bibnamefont{Naldesi}}, \bibnamefont{and}
  \bibinfo{author}{\bibfnamefont{T.}~\bibnamefont{Roscilde}},
  \bibinfo{journal}{Phys. Rev. B} \textbf{\bibinfo{volume}{95}},
  \bibinfo{pages}{245111} (\bibinfo{year}{2017}),
  \urlprefix\url{https://link.aps.org/doi/10.1103/PhysRevB.95.245111}.

\bibitem[{\citenamefont{Monroe et~al.}(2021)\citenamefont{Monroe, Campbell,
  Duan, Gong, Gorshkov, Hess, Islam, Kim, Linke, Pagano
  et~al.}}]{Monroe2021RMP}
\bibinfo{author}{\bibfnamefont{C.}~\bibnamefont{Monroe}},
  \bibinfo{author}{\bibfnamefont{W.~C.} \bibnamefont{Campbell}},
  \bibinfo{author}{\bibfnamefont{L.-M.} \bibnamefont{Duan}},
  \bibinfo{author}{\bibfnamefont{Z.-X.} \bibnamefont{Gong}},
  \bibinfo{author}{\bibfnamefont{A.~V.} \bibnamefont{Gorshkov}},
  \bibinfo{author}{\bibfnamefont{P.~W.} \bibnamefont{Hess}},
  \bibinfo{author}{\bibfnamefont{R.}~\bibnamefont{Islam}},
  \bibinfo{author}{\bibfnamefont{K.}~\bibnamefont{Kim}},
  \bibinfo{author}{\bibfnamefont{N.~M.} \bibnamefont{Linke}},
  \bibinfo{author}{\bibfnamefont{G.}~\bibnamefont{Pagano}},
  \bibnamefont{et~al.}, \bibinfo{journal}{Rev. Mod. Phys.}
  \textbf{\bibinfo{volume}{93}}, \bibinfo{pages}{025001}
  (\bibinfo{year}{2021}),
  \urlprefix\url{https://link.aps.org/doi/10.1103/RevModPhys.93.025001}.

\bibitem[{\citenamefont{Sandvik}(2010)}]{Sandvik2010AIPCP}
\bibinfo{author}{\bibfnamefont{A.~W.} \bibnamefont{Sandvik}},
  \bibinfo{journal}{AIP Conf. Proc.} \textbf{\bibinfo{volume}{1297}},
  \bibinfo{pages}{135} (\bibinfo{year}{2010}),
  \urlprefix\url{https://aip.scitation.org/doi/abs/10.1063/1.3518900}.

\bibitem[{\citenamefont{Agarwal}(2012)}]{Agarwal-book}
\bibinfo{author}{\bibfnamefont{G.~S.} \bibnamefont{Agarwal}},
  \emph{\bibinfo{title}{{Quantum Optics}}} (\bibinfo{publisher}{Cambridge},
  \bibinfo{year}{2012}).

\bibitem[{\citenamefont{Roscilde
  et~al.}(2023{\natexlab{b}})\citenamefont{Roscilde, Comparin, and
  Mezzacapo}}]{Roscildeetal2023b}
\bibinfo{author}{\bibfnamefont{T.}~\bibnamefont{Roscilde}},
  \bibinfo{author}{\bibfnamefont{T.}~\bibnamefont{Comparin}}, \bibnamefont{and}
  \bibinfo{author}{\bibfnamefont{F.}~\bibnamefont{Mezzacapo}},
  \bibinfo{journal}{Phys. Rev. B} \textbf{\bibinfo{volume}{108}},
  \bibinfo{pages}{155130} (\bibinfo{year}{2023}{\natexlab{b}}),
  \urlprefix\url{https://link.aps.org/doi/10.1103/PhysRevB.108.155130}.

\end{thebibliography}

\end{document}